\documentclass[%
 reprint,
superscriptaddress,
 amsmath,amssymb,
 aps,
prl,
]{revtex4-1}

\usepackage{graphicx}
\usepackage{dcolumn}
\usepackage{bm}
\usepackage[colorlinks=true,urlcolor=magenta,linkcolor=magenta,citecolor=cyan]{hyperref}
\usepackage{verbatim}
\usepackage[utf8]{inputenc}
\usepackage{layouts}
\usepackage{xcolor}
\usepackage{siunitx}
\usepackage{gensymb} 

\begin{document}

\title{Tuning the Optical Properties of a MoSe$_2$ Monolayer\\ Using Nanoscale Plasmonic Antennas}

\author{Marko M. Petri\'{c}}
\email{Marko.Petric@wsi.tum.de}
\affiliation{Walter Schottky Institut, Department of Electrical and Computer Engineering and MCQST, Technische Universit\"{a}t M\"{u}nchen, Am Coulombwall 4, 85748 Garching, Germany}

\author{Malte Kremser}
\affiliation{Walter Schottky Institut, Physik-Department and MCQST, Technische Universit\"{a}t M\"{u}nchen, Am Coulombwall 4, 85748 Garching, Germany}

\author{Matteo Barbone}
\affiliation{Walter Schottky Institut, Department of Electrical and Computer Engineering and MCQST, Technische Universit\"{a}t M\"{u}nchen, Am Coulombwall 4, 85748 Garching, Germany}

\author{Anna Nolinder}
\affiliation{Walter Schottky Institut, Physik-Department and MCQST, Technische Universit\"{a}t M\"{u}nchen, Am Coulombwall 4, 85748 Garching, Germany}

\author{Anna Lyamkina}
\affiliation{Walter Schottky Institut, Physik-Department and MCQST, Technische Universit\"{a}t M\"{u}nchen, Am Coulombwall 4, 85748 Garching, Germany}

\author{Andreas V. Stier}
\affiliation{Walter Schottky Institut, Physik-Department and MCQST, Technische Universit\"{a}t M\"{u}nchen, Am Coulombwall 4, 85748 Garching, Germany}

\author{Michael Kaniber}
\affiliation{Walter Schottky Institut, Physik-Department and MCQST, Technische Universit\"{a}t M\"{u}nchen, Am Coulombwall 4, 85748 Garching, Germany}

\author{Kai M\"{u}ller}
\affiliation{Walter Schottky Institut, Department of Electrical and Computer Engineering and MCQST, Technische Universit\"{a}t M\"{u}nchen, Am Coulombwall 4, 85748 Garching, Germany}

\author{Jonathan J. Finley}
\email{Jonathan.Finley@wsi.tum.de}
\affiliation{Walter Schottky Institut, Physik-Department and MCQST, Technische Universit\"{a}t M\"{u}nchen, Am Coulombwall 4, 85748 Garching, Germany}

\begin{abstract}
Nanoplasmonic systems combined with optically active two-dimensional materials provide intriguing opportunities to explore and control light--matter interactions at extreme subwavelength length scales approaching the exciton Bohr radius. Here, we present room- and cryogenic-temperature investigations of light--matter interactions between a MoSe$_2$ monolayer and individual lithographically defined gold dipole nanoantennas having sub-10 nm feed gaps. By progressively tuning the nanoantenna size, their dipolar resonance is tuned relative to the A-exciton transition in a proximal MoSe$_2$ monolayer achieving a total tuning of $\sim\SI{130}{\milli\electronvolt}$. Differential reflectance measurements performed on $>$100 structures reveal an apparent avoided crossing between exciton and dipolar mode and an exciton--plasmon coupling constant of $g=\SI{55}{\milli\electronvolt}$, representing $g/(\hbar\omega_X)\geq3\%$ of the transition energy. This places our hybrid system in the intermediate-coupling regime where spectra exhibit a characteristic Fano-like shape, indicative of the interplay between pronounced light--matter coupling and significant damping. We also demonstrate active control of the optical response by varying the polarization of the excitation light to programmably suppress coupling to the dipole mode. We further study the emerging optical signatures of the monolayer localized at dipole nanoantennas at 10 K. Our findings represent a key step towards realizing nonlinear photonic devices based on 2D materials with potential for low-energy and ultrafast performance.
\end{abstract}

\maketitle

\section{Introduction}

Monolayer transition metal dichalcogenides (TMDs) uniquely combine optical properties and phenomena such as strong light--matter interaction \cite{wurstbauer_lightmatter_2017}, large spin--orbit coupling \cite{zhu_giant_2011}, valley-contrasting spin \cite{xiao_coupled_2012}, quantum light emission \cite{tonndorf_single-photon_2015, koperski_single_2015, chakraborty_voltage-controlled_2015, srivastava_optically_2015, he_single_2015, palacios-berraquero_large-scale_2017, branny_deterministic_2017, klein_engineering_2021}, and large exciton binding energies (small Bohr radii) that allow operation at room temperature \cite{chernikov_exciton_2014, wang_colloquium_2018}. These materials are a promising platform for valleytronics \cite{jones_optical_2013, xu_spin_2014}, quantum photonics \cite{aharonovich_solid-state_2016, atature_material_2018}, and the investigation of few- and many-body physics \cite{mak_tightly_2013, barbone_charge-tuneable_2018, li_revealing_2018, ye_efficient_2018, chen_coulomb-bound_2018, hao_neutral_2017, kremser_discrete_2020, klein_controlling_2021} and strongly correlated phases \cite{costanzo_gate-induced_2016, ye_superconducting_2012}. Furthermore, the ability to fabricate heterostructures due to their van der Waals character allows the investigation of dipolar interlayer excitons \cite{geim_van_2013, rivera_observation_2015, seyler_signatures_2019}.

Large excitonic oscillator strengths and their two-dimensional geometry enable efficient and simple integration of TMDs into photonic structures \cite{tonndorf_-chip_2017, youngblood_integration_2016}, forming hybrid systems that support novel regimes of light--matter interactions, such as nonlinear and nonlocal coupling. These interactions can be engineered well below the diffraction limit through an interplay between localized surface plasmon polaritons (LSPPs) and light emitters \cite{sriram_hybridizing_2020, blauth_enhanced_2017, blauth_coupling_2018}, giving access to the extreme length scales of the order of the exciton Bohr radius \cite{stier_exciton_2016, stier_magnetooptics_2018, goryca_revealing_2019}. In the proximity of metallic nanoparticles that host LSSPs, the photonic density of modes available to a light emitter can be strongly enhanced in two different light--matter coupling regimes: weak-coupling, for which the light--matter coupling is perturbative, and strong-coupling for which it becomes coherent \cite{baranov_novel_2018, pelton_strong_2019}. Weakly coupled systems exhibit modified spontaneous emission rates of the emitter, physics that is captured by the Purcell effect \cite{akselrod_probing_2014}. In strongly coupled systems on the other hand, polaritons having mixed light and matter character emerge \cite{chikkaraddy_single-molecule_2016}. In between, the intermediate-coupling regime displays Fano-shaped scattering spectra, indicative of the interplay between pronounced light--matter coupling and significant damping \cite{miroshnichenko_fano_2010, lee_fano_2015, abid_temperature-dependent_2017, wang_tunable_2018, sun_light-emitting_2018}. 
Recently, plasmonic nanocavities realized using high-quality chemically synthesized metallic nanoparticles have been shown to result in strong and weak light--matter coupling with free excitons in TMDs \cite{kleemann_strong-coupling_2017, wen_room-temperature_2017, zheng_manipulating_2017, geisler_single-crystalline_2019, qin_revealing_2020}. However, the relative positioning of different nanophotonic elements is difficult to achieve using such approaches, thereby hindering routes toward integrated technologies. Strong coupling was also demonstrated via coupling emitters to the delocalized collective modes of metallic nanoparticle arrays \cite{lee_fano_2015, wang_coherent_2016, liu_strong_2016}. 

Lithographically defined nanoresonators that host localized plasmonic modes, such as dielectric \cite{sortino_enhanced_2019} and plasmonic nanoscale antennas \cite{yan_strong_2020}, provide maximum flexibility in design, relative position, and levels of integration.  Furthermore, they can be individually optically probed to peer through ensemble broadening and access the light--matter coupling at the level of a few excitons. Besides coupling to free excitons in such structures, the emission rate of localized emitters can also be strongly enhanced \cite{luo_deterministic_2018, cai_radiative_2018}, making them highly relevant for photonic quantum technologies.

We demonstrate tunable light--matter interaction in a $\mathrm{MoSe_2}$ monolayer proximal to lithographically defined plasmonic antennas having arm sizes centered around $90$ nm and $\leq10$ nm feed gaps. We tune the strength of the light--matter coupling via (i) control of nanoantenna design to spectrally detune the dipolar mode from the excitonic transition and (ii)  polarization control to excite a superposition of single nanoparticle plasmon modes and coupled dipolar modes. First, we designed an array of $\mathrm{MoSe_2}$-coupled dipole nanoantennas using finite-difference time-domain (FDTD) calculations. All antennas were widely separated to allow them to be individually optically addressed. A large number ($\geq100$) of nanoantennas were probed using differential reflectance spectroscopy, each having different exciton--plasmon detunings. An apparent avoided crossing is observed between exciton and the dipolar mode of the antenna, indicative of strong coupling. However, careful analysis using a coupled mode model shows that our system falls into the intermediate-coupling regime. We demonstrate active control of the coupling by switching the hot-spot within the nanoantenna feed gap on and off via the incident excitation polarization. Finally, in low-temperature photoluminescence spectroscopy we observe a redshift of free excitons and the emergence of localized excitons at the position of nanoresonators with a polarization that reflects the nanoscale photonic mode in the nanoantenna. The use of top-down nanolithography to define photonic structures, deterministic polarization control, and site-selective localization of excitons within the nanoscale feed gap set a solid foundation for exploring novel regimes of light--matter coupling such as collective superradiance \cite{temnov_superradiance_2005} and nonlocal light--matter interactions that go beyond the dipole approximation \cite{stobbe_spontaneous_2012}.

\section{Results and Discussion}

\begin{figure*}[!ht]
\includegraphics[width=1\textwidth]{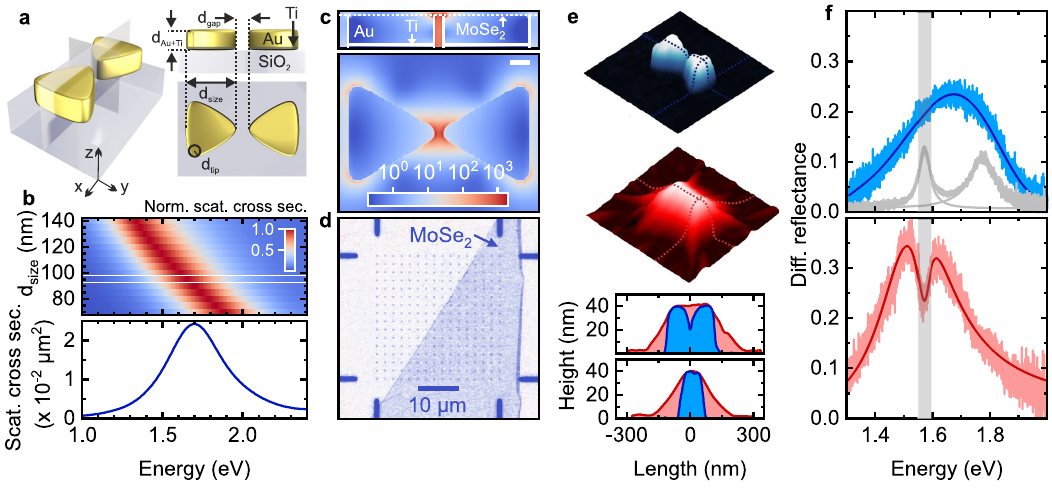}
\caption{\textbf{Geometry and optical response of the $\mathrm{MoSe_2}$-coupled dipole nanoantenna.} \textbf{(a)} Schematic representation of a dipole nanoantenna with highlighted geometrical parameters. \textbf{(b)} Calculated normalized scattering cross section data for dipole nanoantennas with different sizes (top panel) and a highlighted scattering cross section spectrum for $d_{\mathrm{size}} = \SI{90}{\nano\meter}$ (bottom panel) obtained from FDTD calculations. \textbf{(c)} Side-view (top) and top-view (bottom) of the spatial distribution of light intensity enhancement of a $d_{\mathrm{size}} = \SI{90}{\nano\meter}$ sized $\mathrm{MoSe_2}$-coupled dipole nanoantenna calculated via FDTD methods. The dotted line in the top panel represents the $\mathrm{MoSe_2}$ monolayer. Scale bar is $\SI{20}{\nano\meter}$. The side-view contains the long axis of the nanoantenna, while the top-view plane is at the monolayer height. \textbf{(d)} Contrast micrograph of the nanoantenna array covered by the monolayer $\mathrm{MoSe_2}$. The distance between neighboring nanoantennas is $\SI{2}{\micro\meter}$. \textbf{(e)} AFM images of a bare nanoantenna (blue coded) and a $\mathrm{MoSe_2}$-coupled dipole nanoantenna (red coded). Dotted lines are at the positions of height cross sections, depicted at the bottom with respective colors. \textbf{(f)} Differential reflectance spectra recorded from a bare nanoantenna (blue), the bare $\mathrm{MoSe_2}$ monolayer (gray) and a $\mathrm{MoSe_2}$-coupled dipole nanoantenna (red) with corresponding fits.}
\label{fig:Figure1}
\end{figure*}

To achieve the coupling between the $A$-exciton in monolayer $\mathrm{MoSe_2}$ and the nanophotonic resonator, it is necessary to ensure the spectral overlap of the exciton and plasmon resonances, as well as the spatial overlap of the optically active material and the active part of the resonator. Spectral overlap is achieved through engineering the size of the antenna during fabrication. As schematically presented in Figure \ref{fig:Figure1}a, we use dipole bowtie nanoantennas \cite{novotny_antennas_2011, kinkhabwala_large_2009} (see \hyperlink{page.11}{SI} Section \hyperlink{page.11}{S1}), consisting of two lithographically defined triangular gold nanoparticles brought to proximity with a few-nanometer feed gap between them. This geometry results in large localized electric field intensity enhancements due to the near-field dipolar coupling and lightning rod effect \cite{su_interparticle_2003, liao_lightning_1982}, and a single-mode selection through different excitation polarizations \cite{schraml_optical_2014}. To reduce the large parameter space and simplify the spectral control, we nominally fixed all of the parameters to maximize the ratio between quality factor $Q$ and mode volume $V_\mathrm{m}$ as a figure of merit for the strength of the light--matter coupling (see \hyperlink{page.11}{SI} Section \hyperlink{page.11}{S2}) except for the size of a single nanoantenna arm $d_{\mathrm{size}}$, which was used as a free parameter for tuning the plasmonic resonance \cite{schraml_optical_2014}. As the resonator material, we used gold with thickness $d_{\mathrm{Au}} = \SI{35}{\nano\meter}$ on top of the $d_{\mathrm{Ti}} = \SI{5}{\nano\meter}$ titanium layer on the SiO$_2$ substrate. The radius of curvature at the triangle corners is resolution limited to as small as $d_{\mathrm{tip}} = \SI{10}{\nano\meter}$. We fixed the feed gap size to be $d_{\mathrm{gap}} = \SI{10\pm2}{\nano\meter}$, to achieve the lowest $V_\mathrm{m}$ within fabrication limits. We performed FDTD calculations (see \hyperref[sec:Methods]{Methods}) to obtain the dipole resonance frequency as a function of nanoantenna size with the excitation being polarized along the long nanoantenna axis. The results are presented in Figure \ref{fig:Figure1}b (upper panel) and show the simulated scattering cross section as a function of energy and $d_{\mathrm{size}}$. The data show a large redshift, $\geq 0.5$ eV, upon increasing $d_{\mathrm{size}}$ from $70$ to $140$ nm spanning the range of optical activity of MoSe$_2$. To tailor the plasmonic resonance to match the neutral exciton energy in monolayer $\mathrm{MoSe_2}$ at $\SI{1.57}{\electronvolt}$, simulations of bare plasmonic nanoantennas indicate that the optimum size is $d_\mathrm{size} = \SI{110}{\nano\meter}$. However, introducing the MoSe$_2$ monolayer on top of the bowtie antennas was found to result in a large redshift of the resonance by $\SI{\sim200}{\milli\electronvolt}$ (see Figure \hyperlink{page.11}{S5}b) and therefore we compensated by choosing $d_\mathrm{size} = \SI{90}{\nano\meter}$ to achieve good spectral coupling of the nanoantenna mode. Such a large redshift can be explained by the proximity of the high-refractive-index material to the plasmonic hot-spot, demonstrating high sensitivity of plasmonic resonance to the change of the dielectric environment (see Figure \hyperlink{page.11}{S8}). A typical calculated scattering cross section is presented in Figure \ref{fig:Figure1}b (lower panel) for the optimized antenna size. The MoSe$_2$ monolayer was positioned on top of the nanoantennas and FDTD calculations were performed for a $d_{\mathrm{size}} = \SI{90}{\nano\meter}$ titanium--gold nanoantenna covered by $\SI{0.7}{\nano\meter}$ thick monolayer. The resulting side- and top-views of the electromagnetic field intensity enhancement are presented in Figure \ref{fig:Figure1}c, revealing that the strongest field enhancement occurs in the feed gap at the position of the TMD (dotted circle). Maximum intensity enhancements up to $\sim$1$0^3$ are expected for this optimized geometry. A comparison to the similar intensity distribution for the bare nanoantenna in Figure \hyperlink{page.11}{S1} shows that the flake on top "pulls" the electromagnetic hot-spot from the bottom of the feed gap to the top to maximize the spatial coupling with the MoSe$_2$ monolayer.

To realize $\mathrm{MoSe_2}$-coupled dipole nanoantennas, we fabricated arrays of dipole nanoantennas using electron beam lithography (see \hyperref[sec:Methods]{Methods}) with a separation of $\SI{2}{\micro\meter}$ to facilitate optical addressing using confocal microscopy. Simulated scattering cross sections and measured differential reflectance spectra of individual plasmonic nanoantennas (see \hyperlink{page.11}{SI} Section \hyperlink{page.11}{S3}) are in excellent agreement, which allows us to use simulations as a predictive tool to design the desired spectral response of the nanoantenna. The measured $Q$ of the realized structures is $Q = 4.5\pm0.5$. Subsequently, we used dry viscoelastic stamping methods \cite{castellanos-gomez_deterministic_2014} to cover the nanoantenna array with a $\mathrm{MoSe_2}$ monolayer, as depicted in the contrast micrograph in Figure \ref{fig:Figure1}d to finally obtain $\mathrm{MoSe_2}$-coupled dipole nanoantennas. As shown in Figure \ref{fig:Figure1}e, atomic force microscopy (AFM) was used to explore how the monolayer flake covers the nanostructure, where blue-coded and red-coded three-dimensional representations in the figure show a bare nanoantenna and $\mathrm{MoSe_2}$-coupled dipole nanoantenna topographies, respectively. Corresponding cross sections at the dotted lines shown in the figure represent height profiles of the nanoantenna without (blue) and with (red) the TMD monolayer on top, confirming the spatial overlap of the flake and the hot-spot.

To optically characterize our nanostructures (see \hyperref[sec:Methods]{Methods}), and thus probe the underlying light--matter interaction mechanisms, we performed differential reflectance spectroscopy. A typical spectrum recorded from a bare nanoantenna, shown in Figure \ref{fig:Figure1}f (blue), exhibits a broad plasmonic resonance centered at $\SI{\sim1.68}{\electronvolt}$. Because of the asymmetric shape, the spectrum is fitted with a double Lorentzian curve, which may be the result of a small size difference in the two nanoantenna arms. Away from the antennas, the bare flake shows the typical signature of $A$- ($\SI{\sim1.57}{\electronvolt}$) and $B$-excitons ($\SI{\sim1.78}{\electronvolt}$) of the $\mathrm{MoSe_2}$ monolayer at room temperature \cite{li_measurement_2014}, as shown in Figure \ref{fig:Figure1}f (gray). The spectral mismatch between the $A$-exciton and the plasmon resonance of the bare antenna is intentional since, as discussed above, the introduction of a monolayer on top of the antenna redshifts the resonance (see Figure \hyperlink{page.11}{S5}b). The optical response of the $\mathrm{MoSe_2}$-coupled dipole nanoantenna, depicted in red in Figure \ref{fig:Figure1}f, consists of two peaks: the low-energy (LE) peak and high-energy (HE) peak separated by a dip at the $A$-exciton position.


\begin{figure*}[!ht]
\includegraphics[width=1\textwidth]{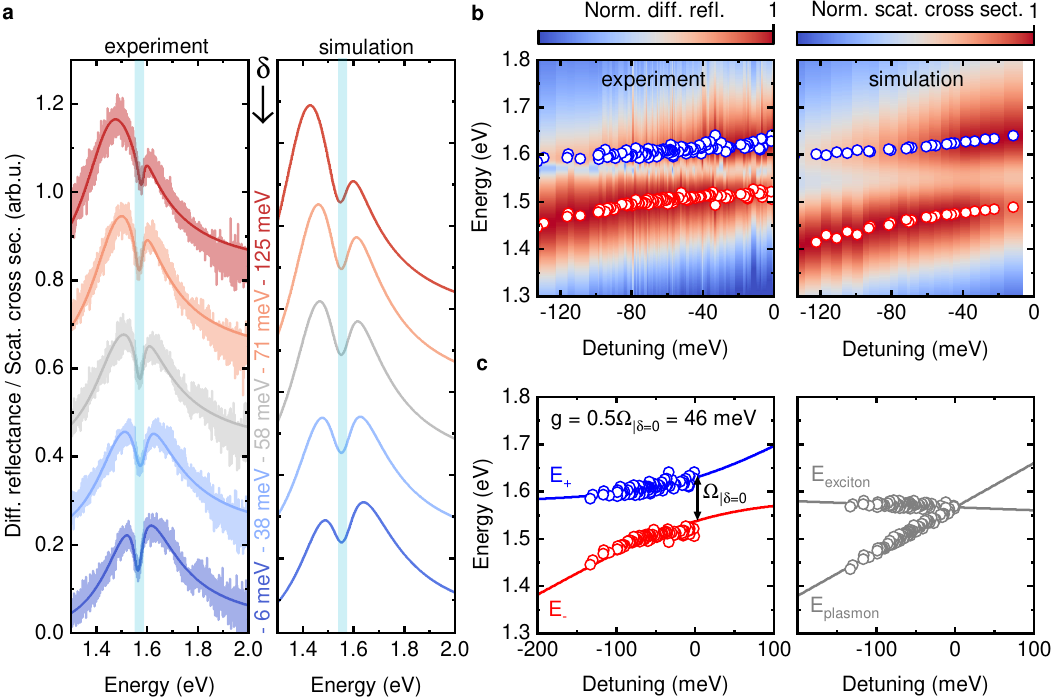}
\caption{\textbf{Probing the nature of the light--matter interaction in $\mathrm{MoSe_2}$-coupled dipole nanoantennas.} \textbf{(a)} Differential reflectance (left) and corresponding calculated scattering cross section spectra (right) of differently detuned nanoantennas in the range from ${\sim}\SI{-130}{\milli\electronvolt}$ up to $\SI{\sim0}{\milli\electronvolt}$. Differential reflectance spectra are accompanied by the COM fits using Eq. (\ref{equ:ABSscat}). The blue band denotes the $A$-exciton spectral position in a $\mathrm{MoSe_2}$ monolayer at room temperature. \textbf{(b)} Normalized differential reflectance (left) and calculated normalized scattering cross section (right) of ${\sim}100$ $\mathrm{MoSe_2}$-coupled dipole nanoantennas ordered by increased detuning of $\delta$. Blue and red circles correspond to the HE- and LE-peaks, respectively. \textbf{(c)} The LE-peak ($E_-$ in red) and HE-peak ($E_+$ in blue) energy  position as a function of the detuning with corresponding fits according to the Eq. (\ref{equ:COM2}) (left panel). Exciton and plasmon energies extracted as fit parameters from Eq. (\ref{equ:ABSscat}) (right panel).} 
\label{fig:Figure2}
\end{figure*}

To understand the nature of the optical response of the hybrid nanoantenna-MoSe$_2$ structure, we analyze the data using a coupled oscillator model (COM) \cite{wu_quantum-dot-induced_2010, pelton_strong_2019}. The COM considers plasmonic and excitonic modes as coupled damped harmonic oscillators driven by an external electric field. Starting from the equations of motion of such a system, we derive the resulting scattering cross section $\sigma_{\mathrm{scat}}$ of the coupled exciton--plasmon system as a function of photon energy $E$. It is given by \cite{wu_quantum-dot-induced_2010}: 
\begin{equation}
\resizebox{.9\hsize}{!}{$
\sigma_{\mathrm{scat}}(E)=AE^4\Bigg|\frac{E^2-E_{\mathrm{A}}^2+iE\Gamma_{\mathrm{A}}}{(E^2-E_{\mathrm{A}}^2+iE\Gamma_{\mathrm{A}})(E^2-E_{\mathrm{plasmon}}^2+iE\Gamma_{\mathrm{plasmon}}) - 4E^2g^2}\Bigg|^2$}
\label{equ:ABSscat}
\end{equation}
where $E_A$ and $E_\mathrm{plasmon}$ are the uncoupled exciton and plasmon resonance energies, respectively, $\Gamma_{\mathrm{A}}$ and $\Gamma_{\mathrm{plasmon}}$ are the corresponding dampings, $g$ is the coupling constant of the two oscillators, and $A$ is an arbitrary scaling factor. For simplicity, we neglected the effect of the $B$-exciton, since it is largely detuned from the plasmonic resonance due to the redshift upon the monolayer transfer and it has weaker oscillator strength compared to the $A$-exciton. In Figure \ref{fig:Figure2}a (left panel), we present the typical spectra recorded from five different $\mathrm{MoSe_2}$-coupled dipole nanoantennas and their respective COM fits that reveal different optical responses depending on the detuning $\delta = E_{\mathrm{plasmon}} - E_{\mathrm{A}}$. Each curve shows a dip in a narrow energy region highlighted in blue and exactly at the A-exciton position of a MoSe$_2$ monolayer. The detuning originates from the slight differences in plasmonic resonances of each dipole nanoantenna due to their spread in size $d_\mathrm{arm} = \SI{90\pm2}{\nano\meter}$ and feed gap size $d_\mathrm{gap} = \SI{10\pm2}{\nano\meter}$, caused by fabrication fluctuations (see \hyperlink{page.11}{SI} Section \hyperlink{page.11}{S4}). The HE-peak asymptotically approaches the exciton energy for increasing negative detuning, while for the LE-peak the same behavior occurs for increasing positive detuning. Complementary to the measurements, the five corresponding calculated spectra are presented in the right panel in Figure \ref{fig:Figure2}a, showing similar behaviour.

In Figure \ref{fig:Figure2}b (left), we plot the normalized differential reflectance spectra fits of ${\sim}100$ $\mathrm{MoSe_2}$-coupled dipole nanoantennas ordered by increasing detuning. The blue and red circles denote HE-peak and LE-peak spectral position, respectively. They clearly show an apparent anticrossing behavior that is very similar to recent reports \cite{kleemann_strong-coupling_2017, wen_room-temperature_2017, zheng_manipulating_2017, geisler_single-crystalline_2019, qin_revealing_2020}. However, we continue to show that the system operates in the intermediate-coupling regime.  The rightmost panel in Figure \ref{fig:Figure2}b shows calculated scattering cross sections of nanoantennas within that range illustrating the excellent agreement between experiment and our modeling. Distributions of the fitting parameters over ${\sim}100$ nanoantennas are presented in \hyperlink{page.11}{SI} Section \hyperlink{page.11}{S5}. To quantify the coupling strength, we again used the COM where the eigenenergies $E_{\pm}$ of the coupled system Hamiltonian are given by the equation:
\begin{equation}
E_{\pm} = \frac{1}{2}(E_{\mathrm{A}} + E_{\mathrm{plasmon}}) \pm \sqrt{g^2 + \frac{1}{4}\delta^2.}
\label{equ:COM2}
\end{equation}
The exciton and plasmon damping terms are neglected under the assumption $\Gamma_{\mathrm{A}}\ll E_{\mathrm{A}}$ and $\Gamma_{\mathrm{plasmon}}\ll E_{\mathrm{plasmon}}$ \cite{zheng_manipulating_2017, chikkaraddy_single-molecule_2016, wen_room-temperature_2017}. The upper and lower anticrossing branches are shown in blue and red in the leftmost panel in Figure \ref{fig:Figure2}c and correspond to $E_+$ (energy of HE-peak) and $E_-$ (energy of LE-peak), respectively. We obtain the coupling constant $g = \frac{1}{2}\Omega_{|\delta = 0} = \SI{46}{\milli\electronvolt}$, where $\Omega_{|\delta = 0}$ is the splitting at zero detuning, as shown in Figure \ref{fig:Figure2}c. This value represents the rate the energy is exchanged between the two interacting systems having predominant light and matter character, respectively. For our system, we obtain $g < g_c$, placing it clearly into the intermediate-coupling regime close to strong-coupling regime \cite{pelton_strong_2019, sun_light-emitting_2018}. Here, $g_c = \frac{1}{2}(\Gamma_{\mathrm{plasmon}} + \Gamma_{\mathrm{A}}) = \SI{90}{\milli\electronvolt}$ is the critical coupling value, defined as the boundary between the two regimes. Consequently, differential reflectance spectra can be interpreted as the result of the interference between the fields associated with exciton and plasmon mode, leading to the characteristic Fano line shape \cite{miroshnichenko_fano_2010, lee_fano_2015, limonov_fano_2017, abid_temperature-dependent_2017, sun_light-emitting_2018}. The right panel in Figure \ref{fig:Figure2}c shows exciton and plasmon energies, where gray circles are extracted as fit parameters from the Eq. (\ref{equ:ABSscat}). The exciton branch is centered near the value of $\SI{1.57}{\electronvolt}$ and the plasmon resonant energy has an expected linear behavior. We also examined another array of $\mathrm{MoSe_2}$-coupled dipole nanoantennas that contained dipole nanoantennas of smaller sizes, and thus higher energies of the plasmonic resonance. This enabled us to reach both sides of the detuning curve with respect to the $A$-exciton and we found a similar coupling strength of $g = \SI{55}{\milli\electronvolt}$ (see \hyperlink{page.11}{SI} Section \hyperlink{page.11}{S6}). Larger positive detunings could not be reproducibly achieved due to the limits of the electron-beam lithography fabrication process.


\begin{figure}[!ht]
\includegraphics{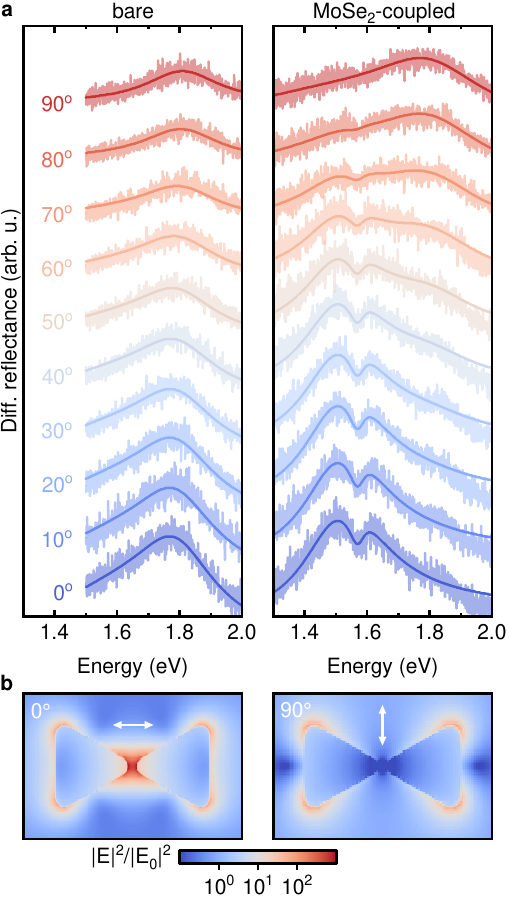}
\caption{\textbf{Active control of the coupling via change of excitation polarization.} \textbf{(a)} Polarization-dependent differential reflectance spectra of a bare nanoantenna fitted by the superposition of the two double Lorentzian functions at 0\degree{} and 90\degree{} (left) and a $\mathrm{MoSe_2}$-coupled dipole nanoantenna fitted by the superposition of the COM for 0\degree{} using Eq. (\ref{equ:ABSscat}) and the double Lorentzian for 90\degree{} (right). \textbf{(b)} The calculated spatial distribution of light intensity enhancement of a bare dipole nanoantenna for excitation light polarized along the long axis, that is, 0\degree{} (left) and perpendicular to the long axis, that is, 90\degree{} (right).}
\label{fig:Figure3}
\end{figure}

We continue to explore active control of the coupling via the polarization of the excitation source. Because of the specific design of dipole nanoantennas, linearly polarized incident light excites a superposition of coupled and decoupled modes upon changing its polarization angle relative to the axis passing between the two triangles through the feed gap \cite{schraml_optical_2014}. To demonstrate this, we performed polarization-dependent differential reflectance measurements on a bare nanoantenna by changing the polarization angle of the broadband laser from 0\degree{} to 90\degree{} (corresponding to the polarization direction along and perpendicular to the long axis of the nanoantenna, respectively). The polarization-dependent spectra in Figure \ref{fig:Figure3}a (left panel) exhibit a clear blueshift of the plasmonic mode by ${\sim}\SI{50}{meV}$ upon turning the polarization from  0\degree{}, parallel to the feed gap, to 90\degree{}, perpendicular to the feed gap. By fitting the 0\degree{} and 90\degree{} differential reflectance spectra that correspond to two different modes using double Lorentzian to take into account the asymmetry in triangle sizes, we obtain the spectra for maximum and minimum nanoparticle coupling, respectively. Fits that correspond to the angles between 0\degree{} and 90\degree{}  were represented as a linear combination of the two distinct modes. Excellent agreement is obtained between experiment and theory.  The reason for the observed blueshift can be readily understood from the calculated spatial distribution of normalized light intensity of such a nanoantenna. This is presented in Figure \ref{fig:Figure3}b for 0\degree{} and 90\degree{} excitation, respectively. Upon exciting at 0\degree{}, along the long axis of the antenna, the dipolar modes of the two nanoparticles oscillate in phase and couple to each other, thereby, lowering the resonance energy. In contrast, by exciting close to 90\degree{}  the two modes are fully decoupled. This results in \textit{higher} energy of the plasmonic resonance and the observed blueshift. Importantly, we note that the coupled-plasmon mode exhibits an electromagnetic hot-spot in the feed gap of the nanoantenna that can be progressively turned off by rotating the polarization to the 90\degree{} configuration to address the decoupled plasmon mode. 

To demonstrate the active coupling control granted by this property, we conducted polarization-dependent differential reflectance measurements on a $\mathrm{MoSe_2}$-covered dipole nanoantenna. Typical data obtained from these measurements are presented in the right panel in Figure \ref{fig:Figure3}a. At 0\degree{}, the hot-spot in the feed gap was turned on and we observed the previously explained Fano-shaped curve. The COM model provides a very good fit of the data. At 90\degree{}, the hot-spot is turned off since only single (uncoupled) nanoparticle plasmon modes are excited, and as a result excitons are no longer coupled to the plasmonic resonator. Therefore, instead of the COM we fit the data using a double Lorentzian function. For intermediate angles, the linear superposition of the two models provides an excellent match to the measured differential reflectance spectra. Furthermore, the blue-shifted plasmonic mode for 90\degree{} excitation occurs at the position resonant with the $B$-exciton. However, no interference between the nanophotonic resonator mode and $B$-exciton can be seen. Therefore, there is no dip at the position of $\SI{1.78}{\electronvolt}$ due to the absence of the hot-spot in the feed gap. This important observation confirms that coupling occurs within the hot-spot of the antenna within the feed gap and not at the edges for example. Therefore, we switch on and off the hotspot and, thereby, the coupling between the $A$-exciton and the plasmonic resonator. Combining this insight with the AFM measurements performed on several flakes (see Figure \hyperlink{page.11}{S7}), we conclude that the coupling is dominated by the nanoantenna design and not by random flake draping, as the monolayer covers the hot-spot in a reproducible fashion.


\begin{figure}[!ht]
\includegraphics{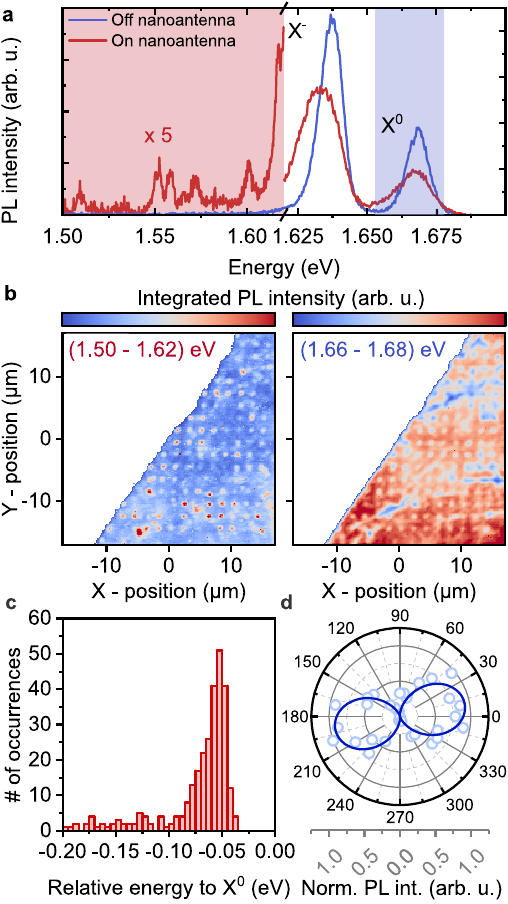}
\caption{\textbf{Modification of $\mathrm{MoSe}_2$ monolayer photoluminescence (PL) at $\SI{10}{\kelvin}$}. \textbf{(a)} Exemplary PL spectra of the $\mathrm{MoSe}_2$ monolayer on a nanoantenna (red) and on the flat SiO$_2$ substrate (blue). $X^0$ and $X^-$ are exciton and negatively charged trion. Blue and red band correspond to the spectral regions $(1.66-1.68)$ $\SI{}{\electronvolt}$ and $(1.50-1.62)$ $\SI{}{\electronvolt}$, respectively. \textbf{(b)} Integrated PL intensity maps with the integration region corresponding to specified spectral regions. \textbf{(c)} Spectral distribution of the localized optical transitions relative to the $\mathrm{X}^0$. \textbf{(d)} PL intensity of an examplary localized optical transition (at $\sim$\SI{1.57}{\electronvolt}) as a function of the excitation polarization angle.}
\label{fig:Figure4}
\end{figure}

Finally, we present investigations of the impact of the nanoantenna on the low-temperature photoluminescence recorded from the MoSe$_2$ monolayer. Figure \ref{fig:Figure4}a shows a typical photoluminescence spectrum recorded at $\SI{10}{\kelvin}$ with $\SI{633}{\nano\meter}$ excitation laser. The data reveal that radiative electron--hole recombination in a semiconducting $\mathrm{MoSe_2}$ monolayer is strongly modified at the position of a $\mathrm{MoSe_2}$-coupled dipole nanoantenna (red spectrum) compared to the flat monolayer region (blue spectrum). On the nanostructure, exciton and trion peak exhibit a redshift and a clear intensity reduction. This shift is most likely due to the gold beneath the MoSe$_2$ that results in locally enhanced screening \cite{rosner_two-dimensional_2016, stier_probing_2016}, bandgap renormalization \cite{ugeda_giant_2014}, and potentially also strain \cite{schmidt_reversible_2016}. The right panel in Figure \ref{fig:Figure4}b shows the integrated photoluminescence intensity for the spectral region $(1.66-1.68)$ $\SI{}{\electronvolt}$ scanned over the whole sample, corresponding to the blue band in Figure \ref{fig:Figure4}a. The integrated intensity consistently shows a dip at the position of nanostructures. The spectrum recorded on the nanoantenna in Figure \ref{fig:Figure4}a reveals several PL features which are red-shifted from the $X^0$ transition. We plot the integrated emission in the spectral range $(1.50-1.62)$ $\SI{}{\electronvolt}$  in the left panel of Figure \ref{fig:Figure4}b and find that the optical transitions in this spectral region are strictly localized at the positions of the nanostructures \cite{branny_discrete_2016, yu_site-controlled_2021}. These transitions possibly stem from locally trapped excitons, whose origin is often attributed to defects \cite{klein_site-selectively_2019}, strain fields \cite{schmidt_reversible_2016}, and the gradient in the dielectric environment \cite{rosner_two-dimensional_2016}. Their occurrences are depicted by the histogram in Figure \ref{fig:Figure4}c, which shows that most transitions occur spectrally close to the low-energy trion tail. We further observed the transitions at the position of dipole nanoantennas and spectrally close to the plasmonic resonance that show strong polarization response in the long-axes direction of the nanostructure as depicted in Figure \ref{fig:Figure4}d for the transition occurring at $\sim$\SI{1.57}{\electronvolt}, which indicates the adaptation of the emission properties to the nanoantenna.


\section{Conclusion}

In summary, we reported optical studies of light--matter coupling between a $\mathrm{MoSe_2}$ monolayer and proximal plasmonic dipole nanoantennas. Despite observing a clear anticrossing in reflectance spectra, we showed that the system operates in the intermediate-coupling regime and that it can be site-selectively accessed and actively controlled at room temperature. Furthermore, these structures exhibit a modified $\mathrm{MoSe_2}$ monolayer photoluminescence that shows a redshift of free excitons and the emergence of trapped excitons. Our methods enable further engineering of the hybrid TMD-metal platform to deterministically position arrays of weakly and strongly coupled systems of a nanoantenna and a single-photon emitter in TMDs. Achieving systems that operate deeper into the strong-coupling regime would require engineering of higher $Q$ \cite{chen_high-_2018} plasmonic structures, for example, using novel plasmonic materials or dielectrics. By site selectively inducing one or more color centers using, for example, local He-ion irradiation \cite{klein_site-selectively_2019}, it may be possible to study novel regimes of few-emitter cavity quantum electrodynamics such as Dicke Superradiance or collective effects, while controlling hybrid exciton--polariton states via hot-spot switching. Additional control could be gained by introduction of a static electric field within the feed gap \cite{kern_electrically_2015}.


\section{Methods}
\label{sec:Methods}

\subsection{FDTD Calculations}
Calculations were performed using 3D Maxwell's equations solver Lumerical FDTD Solutions. Bowtie dipole nanoantennas were modeled as triangles with rounded edges. Except for nanoantenna size, all geometrical parameters were fixed as noted in the main text and \hyperlink{page.11}{SI} Section \hyperlink{page.11}{S2}. For the optical constants of gold, we used values reported by Johnson and Christy \cite{johnson_optical_1972}. For a monolayer MoSe$_2$ at room temperature, the data was extracted from Y. Li \textit{et al.} \cite{li_measurement_2014}. A fine mesh in proximity of the dipole nanoantenna was used with a Yee-cell of $1\times1\times\SI{1}{\nano\meter^3}$. Perfectly matched layer (PML) boundary conditions were applied. To speed up the simulation time, periodic boundary conditions were used to exploit the symmetrical character of our structures. Total-field scattered-field (TFSF) was used as the excitation source, which facilitated the calculation of scattered and absorbed light for different spectral ranges and polarization angles.

\subsection{Sample Fabrication}
The plasmonic dipole nanoantennas were written via electron-beam lithography with an eLine system by Raith. Polymethylmethacrylat (PMMA) 950K, AR-P 679.02, ALLRESIST was spin-coated onto the SiO$_2$ substrate, ramping up from 0 to 4000 rpm for one second, followed by spinning at 4000 rpm for 40 seconds. The PMMA was baked at \SI{170}{\celsius} for 5 minutes. AR-PC 5090 Electra 92 was used to avoid the charging effects (rotation speed 2000 rpm, initial ramp 2000 rpm, spinning time 60 s, \SI{90}{\degree} baking temperature). 30 kV acceleration voltage, \SI{10}{\micro\meter} aperture and 400-\SI{800}{\micro\coulomb/\centi\meter^2} doses were used. Nanoantennas were positioned  \SI{2}{\micro\meter} from each other to avoid coupling of neighboring nanoantennas. After writing, the conductive coating was removed using distilled water and the structures were developed applying Methylisobutylketone (MIBK) and isopropanol 1:3 solution for 45 s. Evaporation of 5 nm titanium and 35 nm gold layer and subsequent lift-off by acetone was performed to obtain the final nanoantenna structures. In the second step, an all-dry viscoelastic (Polydimethylsiloxane) stamping method \cite{castellanos-gomez_deterministic_2014} was used to position and transfer a mechanically exfoliated MoSe$_2$ monolayer on top of plasmonic nanostructures, to finally obtain MoSe$_2$-coupled dipole nanoantennas. The monolayer was identified using atomic force microscopy, contrast in an optical microscope, and differential reflectance measurements. 

\subsection{Optical Measurements}
Differential reflectance and photoluminescence measurements were performed using a self-built confocal microscope. For differential reflectance measurements, a broad-band supercontinuum laser beam (Fianium WhiteLase) was focused on the sample by an objective with N.A. = 0.9 to a diffraction-limited spot. Differential reflectance spectra $\Delta R(E)/R(E)$ were obtained by $(I_{ON}(E) - I_{OFF}(E))/I_{OFF}(E)$, where $I_{ON}(E)$ and $I_{OFF}(E)$ represent the number of counts from reflected light when the excitation laser spot is on a nanoantenna and on a substrate/monolayer, respectively. The light was dispersed by a 150 lines/mm grating onto a charge-coupled device (Horiba). Photoluminescence measurements were performed at \SI{10}{\kelvin} using He-flow cryostat (Cryovac). The excitation He-Ne (\SI{633}{\nano\meter}) laser spot was focused on the sample by an objective with N.A. = 0.75. The used grating was 600 lines/mm.


\section{Author Contributions}

M.M.P. did the FDTD calculations and fabricated the samples. M.M.P. and A.N. constructed the differential reflectance measurement setup and performed differential reflectance measurements and data analysis. M.M.P. and M.Kr. performed PL measurements and data analysis. M.Ka., K.M., and J.J.F. conceived and managed the project. All authors participated in the discussion of the results and the writing of the manuscript.


\section{Acknowledgements}

We thank M. Altzschner, J. Wierzbowski, and M. Blauth for their technical assistance and scientific discussions. M.M.P. acknowledges TUM International Graduate School of Science and Engineering (IGSSE). M.Kr. acknowledges support from the International Max Planck Research School for Quantum Science and Technology (IMPRS-QST). J.J.F. gratefully acknowledges the German Science Foundation (DFG) for financial support via the Clusters of Excellence e.Conversion (EXC 2089), MCQST (EXC 2111), as well as the individual projects DI 2013/5-1 and FI 947/8-1. M.B. and A.L. acknowledge support from the Alexander von Humboldt Foundation. K.M. acknowledges support from the Bavarian Academy of Sciences and Humanities.

\bibliography{main}

\begin{thebibliography}{79}%
\makeatletter
\providecommand \@ifxundefined [1]{%
 \@ifx{#1\undefined}
}%
\providecommand \@ifnum [1]{%
 \ifnum #1\expandafter \@firstoftwo
 \else \expandafter \@secondoftwo
 \fi
}%
\providecommand \@ifx [1]{%
 \ifx #1\expandafter \@firstoftwo
 \else \expandafter \@secondoftwo
 \fi
}%
\providecommand \natexlab [1]{#1}%
\providecommand \enquote  [1]{``#1''}%
\providecommand \bibnamefont  [1]{#1}%
\providecommand \bibfnamefont [1]{#1}%
\providecommand \citenamefont [1]{#1}%
\providecommand \href@noop [0]{\@secondoftwo}%
\providecommand \href [0]{\begingroup \@sanitize@url \@href}%
\providecommand \@href[1]{\@@startlink{#1}\@@href}%
\providecommand \@@href[1]{\endgroup#1\@@endlink}%
\providecommand \@sanitize@url [0]{\catcode `\\12\catcode `\$12\catcode
  `\&12\catcode `\#12\catcode `\^12\catcode `\_12\catcode `\%12\relax}%
\providecommand \@@startlink[1]{}%
\providecommand \@@endlink[0]{}%
\providecommand \url  [0]{\begingroup\@sanitize@url \@url }%
\providecommand \@url [1]{\endgroup\@href {#1}{\urlprefix }}%
\providecommand \urlprefix  [0]{URL }%
\providecommand \Eprint [0]{\href }%
\providecommand \doibase [0]{http://dx.doi.org/}%
\providecommand \selectlanguage [0]{\@gobble}%
\providecommand \bibinfo  [0]{\@secondoftwo}%
\providecommand \bibfield  [0]{\@secondoftwo}%
\providecommand \translation [1]{[#1]}%
\providecommand \BibitemOpen [0]{}%
\providecommand \bibitemStop [0]{}%
\providecommand \bibitemNoStop [0]{.\EOS\space}%
\providecommand \EOS [0]{\spacefactor3000\relax}%
\providecommand \BibitemShut  [1]{\csname bibitem#1\endcsname}%
\let\auto@bib@innerbib\@empty
\bibitem [{\citenamefont {Wurstbauer}\ \emph {et~al.}(2017)\citenamefont
  {Wurstbauer}, \citenamefont {Miller}, \citenamefont {Parzinger},\ and\
  \citenamefont {Holleitner}}]{wurstbauer_lightmatter_2017}%
  \BibitemOpen
  \bibfield  {author} {\bibinfo {author} {\bibfnamefont {U.}~\bibnamefont
  {Wurstbauer}}, \bibinfo {author} {\bibfnamefont {B.}~\bibnamefont {Miller}},
  \bibinfo {author} {\bibfnamefont {E.}~\bibnamefont {Parzinger}}, \ and\
  \bibinfo {author} {\bibfnamefont {A.~W.}\ \bibnamefont {Holleitner}},\ }\href
  {\doibase 10.1088/1361-6463/aa5f81} {\bibfield  {journal} {\bibinfo
  {journal} {J. Phys. D: Appl. Phys.}\ }\textbf {\bibinfo {volume} {50}},\
  \bibinfo {pages} {173001} (\bibinfo {year} {2017})}\BibitemShut {NoStop}%
\bibitem [{\citenamefont {Zhu}\ \emph {et~al.}(2011)\citenamefont {Zhu},
  \citenamefont {Cheng},\ and\ \citenamefont
  {Schwingenschl{\"o}gl}}]{zhu_giant_2011}%
  \BibitemOpen
  \bibfield  {author} {\bibinfo {author} {\bibfnamefont {Z.~Y.}\ \bibnamefont
  {Zhu}}, \bibinfo {author} {\bibfnamefont {Y.~C.}\ \bibnamefont {Cheng}}, \
  and\ \bibinfo {author} {\bibfnamefont {U.}~\bibnamefont
  {Schwingenschl{\"o}gl}},\ }\href {\doibase 10.1103/PhysRevB.84.153402}
  {\bibfield  {journal} {\bibinfo  {journal} {Phys. Rev. B}\ }\textbf {\bibinfo
  {volume} {84}},\ \bibinfo {pages} {153402} (\bibinfo {year}
  {2011})}\BibitemShut {NoStop}%
\bibitem [{\citenamefont {Xiao}\ \emph {et~al.}(2012)\citenamefont {Xiao},
  \citenamefont {Liu}, \citenamefont {Feng}, \citenamefont {Xu},\ and\
  \citenamefont {Yao}}]{xiao_coupled_2012}%
  \BibitemOpen
  \bibfield  {author} {\bibinfo {author} {\bibfnamefont {D.}~\bibnamefont
  {Xiao}}, \bibinfo {author} {\bibfnamefont {G.-B.}\ \bibnamefont {Liu}},
  \bibinfo {author} {\bibfnamefont {W.}~\bibnamefont {Feng}}, \bibinfo {author}
  {\bibfnamefont {X.}~\bibnamefont {Xu}}, \ and\ \bibinfo {author}
  {\bibfnamefont {W.}~\bibnamefont {Yao}},\ }\href {\doibase
  10.1103/PhysRevLett.108.196802} {\bibfield  {journal} {\bibinfo  {journal}
  {Phys. Rev. Lett.}\ }\textbf {\bibinfo {volume} {108}},\ \bibinfo {pages}
  {196802} (\bibinfo {year} {2012})}\BibitemShut {NoStop}%
\bibitem [{\citenamefont {Tonndorf}\ \emph {et~al.}(2015)\citenamefont
  {Tonndorf}, \citenamefont {Schmidt}, \citenamefont {Schneider}, \citenamefont
  {Kern}, \citenamefont {Buscema}, \citenamefont {Steele}, \citenamefont
  {Castellanos-Gomez}, \citenamefont {van~der Zant}, \citenamefont
  {Michaelis~de Vasconcellos},\ and\ \citenamefont
  {Bratschitsch}}]{tonndorf_single-photon_2015}%
  \BibitemOpen
  \bibfield  {author} {\bibinfo {author} {\bibfnamefont {P.}~\bibnamefont
  {Tonndorf}}, \bibinfo {author} {\bibfnamefont {R.}~\bibnamefont {Schmidt}},
  \bibinfo {author} {\bibfnamefont {R.}~\bibnamefont {Schneider}}, \bibinfo
  {author} {\bibfnamefont {J.}~\bibnamefont {Kern}}, \bibinfo {author}
  {\bibfnamefont {M.}~\bibnamefont {Buscema}}, \bibinfo {author} {\bibfnamefont
  {G.~A.}\ \bibnamefont {Steele}}, \bibinfo {author} {\bibfnamefont
  {A.}~\bibnamefont {Castellanos-Gomez}}, \bibinfo {author} {\bibfnamefont
  {H.~S.~J.}\ \bibnamefont {van~der Zant}}, \bibinfo {author} {\bibfnamefont
  {S.}~\bibnamefont {Michaelis~de Vasconcellos}}, \ and\ \bibinfo {author}
  {\bibfnamefont {R.}~\bibnamefont {Bratschitsch}},\ }\href {\doibase
  10.1364/OPTICA.2.000347} {\bibfield  {journal} {\bibinfo  {journal} {Optica}\
  }\textbf {\bibinfo {volume} {2}},\ \bibinfo {pages} {347} (\bibinfo {year}
  {2015})}\BibitemShut {NoStop}%
\bibitem [{\citenamefont {Koperski}\ \emph {et~al.}(2015)\citenamefont
  {Koperski}, \citenamefont {Nogajewski}, \citenamefont {Arora}, \citenamefont
  {Cherkez}, \citenamefont {Mallet}, \citenamefont {Veuillen}, \citenamefont
  {Marcus}, \citenamefont {Kossacki},\ and\ \citenamefont
  {Potemski}}]{koperski_single_2015}%
  \BibitemOpen
  \bibfield  {author} {\bibinfo {author} {\bibfnamefont {M.}~\bibnamefont
  {Koperski}}, \bibinfo {author} {\bibfnamefont {K.}~\bibnamefont
  {Nogajewski}}, \bibinfo {author} {\bibfnamefont {A.}~\bibnamefont {Arora}},
  \bibinfo {author} {\bibfnamefont {V.}~\bibnamefont {Cherkez}}, \bibinfo
  {author} {\bibfnamefont {P.}~\bibnamefont {Mallet}}, \bibinfo {author}
  {\bibfnamefont {J.-Y.}\ \bibnamefont {Veuillen}}, \bibinfo {author}
  {\bibfnamefont {J.}~\bibnamefont {Marcus}}, \bibinfo {author} {\bibfnamefont
  {P.}~\bibnamefont {Kossacki}}, \ and\ \bibinfo {author} {\bibfnamefont
  {M.}~\bibnamefont {Potemski}},\ }\href {\doibase 10.1038/nnano.2015.67}
  {\bibfield  {journal} {\bibinfo  {journal} {Nature Nanotech}\ }\textbf
  {\bibinfo {volume} {10}},\ \bibinfo {pages} {503} (\bibinfo {year}
  {2015})}\BibitemShut {NoStop}%
\bibitem [{\citenamefont {Chakraborty}\ \emph {et~al.}(2015)\citenamefont
  {Chakraborty}, \citenamefont {Kinnischtzke}, \citenamefont {Goodfellow},
  \citenamefont {Beams},\ and\ \citenamefont
  {Vamivakas}}]{chakraborty_voltage-controlled_2015}%
  \BibitemOpen
  \bibfield  {author} {\bibinfo {author} {\bibfnamefont {C.}~\bibnamefont
  {Chakraborty}}, \bibinfo {author} {\bibfnamefont {L.}~\bibnamefont
  {Kinnischtzke}}, \bibinfo {author} {\bibfnamefont {K.~M.}\ \bibnamefont
  {Goodfellow}}, \bibinfo {author} {\bibfnamefont {R.}~\bibnamefont {Beams}}, \
  and\ \bibinfo {author} {\bibfnamefont {A.~N.}\ \bibnamefont {Vamivakas}},\
  }\href {\doibase 10.1038/nnano.2015.79} {\bibfield  {journal} {\bibinfo
  {journal} {Nature Nanotech}\ }\textbf {\bibinfo {volume} {10}},\ \bibinfo
  {pages} {507} (\bibinfo {year} {2015})}\BibitemShut {NoStop}%
\bibitem [{\citenamefont {Srivastava}\ \emph {et~al.}(2015)\citenamefont
  {Srivastava}, \citenamefont {Sidler}, \citenamefont {Allain}, \citenamefont
  {Lembke}, \citenamefont {Kis},\ and\ \citenamefont {Imamo{\u
  g}lu}}]{srivastava_optically_2015}%
  \BibitemOpen
  \bibfield  {author} {\bibinfo {author} {\bibfnamefont {A.}~\bibnamefont
  {Srivastava}}, \bibinfo {author} {\bibfnamefont {M.}~\bibnamefont {Sidler}},
  \bibinfo {author} {\bibfnamefont {A.~V.}\ \bibnamefont {Allain}}, \bibinfo
  {author} {\bibfnamefont {D.~S.}\ \bibnamefont {Lembke}}, \bibinfo {author}
  {\bibfnamefont {A.}~\bibnamefont {Kis}}, \ and\ \bibinfo {author}
  {\bibfnamefont {A.}~\bibnamefont {Imamo{\u g}lu}},\ }\href {\doibase
  10.1038/nnano.2015.60} {\bibfield  {journal} {\bibinfo  {journal} {Nature
  Nanotech}\ }\textbf {\bibinfo {volume} {10}},\ \bibinfo {pages} {491}
  (\bibinfo {year} {2015})}\BibitemShut {NoStop}%
\bibitem [{\citenamefont {He}\ \emph {et~al.}(2015)\citenamefont {He},
  \citenamefont {Clark}, \citenamefont {Schaibley}, \citenamefont {He},
  \citenamefont {Chen}, \citenamefont {Wei}, \citenamefont {Ding},
  \citenamefont {Zhang}, \citenamefont {Yao}, \citenamefont {Xu}, \citenamefont
  {Lu},\ and\ \citenamefont {Pan}}]{he_single_2015}%
  \BibitemOpen
  \bibfield  {author} {\bibinfo {author} {\bibfnamefont {Y.-M.}\ \bibnamefont
  {He}}, \bibinfo {author} {\bibfnamefont {G.}~\bibnamefont {Clark}}, \bibinfo
  {author} {\bibfnamefont {J.~R.}\ \bibnamefont {Schaibley}}, \bibinfo {author}
  {\bibfnamefont {Y.}~\bibnamefont {He}}, \bibinfo {author} {\bibfnamefont
  {M.-C.}\ \bibnamefont {Chen}}, \bibinfo {author} {\bibfnamefont {Y.-J.}\
  \bibnamefont {Wei}}, \bibinfo {author} {\bibfnamefont {X.}~\bibnamefont
  {Ding}}, \bibinfo {author} {\bibfnamefont {Q.}~\bibnamefont {Zhang}},
  \bibinfo {author} {\bibfnamefont {W.}~\bibnamefont {Yao}}, \bibinfo {author}
  {\bibfnamefont {X.}~\bibnamefont {Xu}}, \bibinfo {author} {\bibfnamefont
  {C.-Y.}\ \bibnamefont {Lu}}, \ and\ \bibinfo {author} {\bibfnamefont {J.-W.}\
  \bibnamefont {Pan}},\ }\href {\doibase 10.1038/nnano.2015.75} {\bibfield
  {journal} {\bibinfo  {journal} {Nature Nanotech}\ }\textbf {\bibinfo {volume}
  {10}},\ \bibinfo {pages} {497} (\bibinfo {year} {2015})}\BibitemShut
  {NoStop}%
\bibitem [{\citenamefont {Palacios-Berraquero}\ \emph
  {et~al.}(2017)\citenamefont {Palacios-Berraquero}, \citenamefont {Kara},
  \citenamefont {Montblanch}, \citenamefont {Barbone}, \citenamefont
  {Latawiec}, \citenamefont {Yoon}, \citenamefont {Ott}, \citenamefont
  {Loncar}, \citenamefont {Ferrari},\ and\ \citenamefont
  {Atat{\"u}re}}]{palacios-berraquero_large-scale_2017}%
  \BibitemOpen
  \bibfield  {author} {\bibinfo {author} {\bibfnamefont {C.}~\bibnamefont
  {Palacios-Berraquero}}, \bibinfo {author} {\bibfnamefont {D.~M.}\
  \bibnamefont {Kara}}, \bibinfo {author} {\bibfnamefont {A.~R.-P.}\
  \bibnamefont {Montblanch}}, \bibinfo {author} {\bibfnamefont
  {M.}~\bibnamefont {Barbone}}, \bibinfo {author} {\bibfnamefont
  {P.}~\bibnamefont {Latawiec}}, \bibinfo {author} {\bibfnamefont
  {D.}~\bibnamefont {Yoon}}, \bibinfo {author} {\bibfnamefont {A.~K.}\
  \bibnamefont {Ott}}, \bibinfo {author} {\bibfnamefont {M.}~\bibnamefont
  {Loncar}}, \bibinfo {author} {\bibfnamefont {A.~C.}\ \bibnamefont {Ferrari}},
  \ and\ \bibinfo {author} {\bibfnamefont {M.}~\bibnamefont {Atat{\"u}re}},\
  }\href {\doibase 10.1038/ncomms15093} {\bibfield  {journal} {\bibinfo
  {journal} {Nat Commun}\ }\textbf {\bibinfo {volume} {8}},\ \bibinfo {pages}
  {15093} (\bibinfo {year} {2017})}\BibitemShut {NoStop}%
\bibitem [{\citenamefont {Branny}\ \emph {et~al.}(2017)\citenamefont {Branny},
  \citenamefont {Kumar}, \citenamefont {Proux},\ and\ \citenamefont
  {Gerardot}}]{branny_deterministic_2017}%
  \BibitemOpen
  \bibfield  {author} {\bibinfo {author} {\bibfnamefont {A.}~\bibnamefont
  {Branny}}, \bibinfo {author} {\bibfnamefont {S.}~\bibnamefont {Kumar}},
  \bibinfo {author} {\bibfnamefont {R.}~\bibnamefont {Proux}}, \ and\ \bibinfo
  {author} {\bibfnamefont {B.~D.}\ \bibnamefont {Gerardot}},\ }\href {\doibase
  10.1038/ncomms15053} {\bibfield  {journal} {\bibinfo  {journal} {Nat Commun}\
  }\textbf {\bibinfo {volume} {8}},\ \bibinfo {pages} {15053} (\bibinfo {year}
  {2017})}\BibitemShut {NoStop}%
\bibitem [{\citenamefont {Klein}\ \emph
  {et~al.}(2021{\natexlab{a}})\citenamefont {Klein}, \citenamefont {Sigl},
  \citenamefont {Gyger}, \citenamefont {Barthelmi}, \citenamefont {Florian},
  \citenamefont {Rey}, \citenamefont {Taniguchi}, \citenamefont {Watanabe},
  \citenamefont {Jahnke}, \citenamefont {Kastl}, \citenamefont {Zwiller},
  \citenamefont {J{\"o}ns}, \citenamefont {M{\"u}ller}, \citenamefont
  {Wurstbauer}, \citenamefont {Finley},\ and\ \citenamefont
  {Holleitner}}]{klein_engineering_2021}%
  \BibitemOpen
  \bibfield  {author} {\bibinfo {author} {\bibfnamefont {J.}~\bibnamefont
  {Klein}}, \bibinfo {author} {\bibfnamefont {L.}~\bibnamefont {Sigl}},
  \bibinfo {author} {\bibfnamefont {S.}~\bibnamefont {Gyger}}, \bibinfo
  {author} {\bibfnamefont {K.}~\bibnamefont {Barthelmi}}, \bibinfo {author}
  {\bibfnamefont {M.}~\bibnamefont {Florian}}, \bibinfo {author} {\bibfnamefont
  {S.}~\bibnamefont {Rey}}, \bibinfo {author} {\bibfnamefont {T.}~\bibnamefont
  {Taniguchi}}, \bibinfo {author} {\bibfnamefont {K.}~\bibnamefont {Watanabe}},
  \bibinfo {author} {\bibfnamefont {F.}~\bibnamefont {Jahnke}}, \bibinfo
  {author} {\bibfnamefont {C.}~\bibnamefont {Kastl}}, \bibinfo {author}
  {\bibfnamefont {V.}~\bibnamefont {Zwiller}}, \bibinfo {author} {\bibfnamefont
  {K.~D.}\ \bibnamefont {J{\"o}ns}}, \bibinfo {author} {\bibfnamefont
  {K.}~\bibnamefont {M{\"u}ller}}, \bibinfo {author} {\bibfnamefont
  {U.}~\bibnamefont {Wurstbauer}}, \bibinfo {author} {\bibfnamefont {J.~J.}\
  \bibnamefont {Finley}}, \ and\ \bibinfo {author} {\bibfnamefont {A.~W.}\
  \bibnamefont {Holleitner}},\ }\href {\doibase 10.1021/acsphotonics.0c01907}
  {\bibfield  {journal} {\bibinfo  {journal} {ACS Photonics}\ }\textbf
  {\bibinfo {volume} {8}},\ \bibinfo {pages} {669} (\bibinfo {year}
  {2021}{\natexlab{a}})}\BibitemShut {NoStop}%
\bibitem [{\citenamefont {Chernikov}\ \emph {et~al.}(2014)\citenamefont
  {Chernikov}, \citenamefont {Berkelbach}, \citenamefont {Hill}, \citenamefont
  {Rigosi}, \citenamefont {Li}, \citenamefont {Aslan}, \citenamefont
  {Reichman}, \citenamefont {Hybertsen},\ and\ \citenamefont
  {Heinz}}]{chernikov_exciton_2014}%
  \BibitemOpen
  \bibfield  {author} {\bibinfo {author} {\bibfnamefont {A.}~\bibnamefont
  {Chernikov}}, \bibinfo {author} {\bibfnamefont {T.~C.}\ \bibnamefont
  {Berkelbach}}, \bibinfo {author} {\bibfnamefont {H.~M.}\ \bibnamefont
  {Hill}}, \bibinfo {author} {\bibfnamefont {A.}~\bibnamefont {Rigosi}},
  \bibinfo {author} {\bibfnamefont {Y.}~\bibnamefont {Li}}, \bibinfo {author}
  {\bibfnamefont {O.~B.}\ \bibnamefont {Aslan}}, \bibinfo {author}
  {\bibfnamefont {D.~R.}\ \bibnamefont {Reichman}}, \bibinfo {author}
  {\bibfnamefont {M.~S.}\ \bibnamefont {Hybertsen}}, \ and\ \bibinfo {author}
  {\bibfnamefont {T.~F.}\ \bibnamefont {Heinz}},\ }\href {\doibase
  10.1103/PhysRevLett.113.076802} {\bibfield  {journal} {\bibinfo  {journal}
  {Phys. Rev. Lett.}\ }\textbf {\bibinfo {volume} {113}},\ \bibinfo {pages}
  {076802} (\bibinfo {year} {2014})}\BibitemShut {NoStop}%
\bibitem [{\citenamefont {Wang}\ \emph
  {et~al.}(2018{\natexlab{a}})\citenamefont {Wang}, \citenamefont {Chernikov},
  \citenamefont {Glazov}, \citenamefont {Heinz}, \citenamefont {Marie},
  \citenamefont {Amand},\ and\ \citenamefont
  {Urbaszek}}]{wang_colloquium_2018}%
  \BibitemOpen
  \bibfield  {author} {\bibinfo {author} {\bibfnamefont {G.}~\bibnamefont
  {Wang}}, \bibinfo {author} {\bibfnamefont {A.}~\bibnamefont {Chernikov}},
  \bibinfo {author} {\bibfnamefont {M.~M.}\ \bibnamefont {Glazov}}, \bibinfo
  {author} {\bibfnamefont {T.~F.}\ \bibnamefont {Heinz}}, \bibinfo {author}
  {\bibfnamefont {X.}~\bibnamefont {Marie}}, \bibinfo {author} {\bibfnamefont
  {T.}~\bibnamefont {Amand}}, \ and\ \bibinfo {author} {\bibfnamefont
  {B.}~\bibnamefont {Urbaszek}},\ }\href {\doibase
  10.1103/RevModPhys.90.021001} {\bibfield  {journal} {\bibinfo  {journal}
  {Rev. Mod. Phys.}\ }\textbf {\bibinfo {volume} {90}},\ \bibinfo {pages}
  {021001} (\bibinfo {year} {2018}{\natexlab{a}})}\BibitemShut {NoStop}%
\bibitem [{\citenamefont {Jones}\ \emph {et~al.}(2013)\citenamefont {Jones},
  \citenamefont {Yu}, \citenamefont {Ghimire}, \citenamefont {Wu},
  \citenamefont {Aivazian}, \citenamefont {Ross}, \citenamefont {Zhao},
  \citenamefont {Yan}, \citenamefont {Mandrus}, \citenamefont {Xiao},
  \citenamefont {Yao},\ and\ \citenamefont {Xu}}]{jones_optical_2013}%
  \BibitemOpen
  \bibfield  {author} {\bibinfo {author} {\bibfnamefont {A.~M.}\ \bibnamefont
  {Jones}}, \bibinfo {author} {\bibfnamefont {H.}~\bibnamefont {Yu}}, \bibinfo
  {author} {\bibfnamefont {N.~J.}\ \bibnamefont {Ghimire}}, \bibinfo {author}
  {\bibfnamefont {S.}~\bibnamefont {Wu}}, \bibinfo {author} {\bibfnamefont
  {G.}~\bibnamefont {Aivazian}}, \bibinfo {author} {\bibfnamefont {J.~S.}\
  \bibnamefont {Ross}}, \bibinfo {author} {\bibfnamefont {B.}~\bibnamefont
  {Zhao}}, \bibinfo {author} {\bibfnamefont {J.}~\bibnamefont {Yan}}, \bibinfo
  {author} {\bibfnamefont {D.~G.}\ \bibnamefont {Mandrus}}, \bibinfo {author}
  {\bibfnamefont {D.}~\bibnamefont {Xiao}}, \bibinfo {author} {\bibfnamefont
  {W.}~\bibnamefont {Yao}}, \ and\ \bibinfo {author} {\bibfnamefont
  {X.}~\bibnamefont {Xu}},\ }\href {\doibase 10.1038/nnano.2013.151} {\bibfield
   {journal} {\bibinfo  {journal} {Nature Nanotech}\ }\textbf {\bibinfo
  {volume} {8}},\ \bibinfo {pages} {634} (\bibinfo {year} {2013})}\BibitemShut
  {NoStop}%
\bibitem [{\citenamefont {Xu}\ \emph {et~al.}(2014)\citenamefont {Xu},
  \citenamefont {Yao}, \citenamefont {Xiao},\ and\ \citenamefont
  {Heinz}}]{xu_spin_2014}%
  \BibitemOpen
  \bibfield  {author} {\bibinfo {author} {\bibfnamefont {X.}~\bibnamefont
  {Xu}}, \bibinfo {author} {\bibfnamefont {W.}~\bibnamefont {Yao}}, \bibinfo
  {author} {\bibfnamefont {D.}~\bibnamefont {Xiao}}, \ and\ \bibinfo {author}
  {\bibfnamefont {T.~F.}\ \bibnamefont {Heinz}},\ }\href {\doibase
  10.1038/nphys2942} {\bibfield  {journal} {\bibinfo  {journal} {Nature Phys}\
  }\textbf {\bibinfo {volume} {10}},\ \bibinfo {pages} {343} (\bibinfo {year}
  {2014})}\BibitemShut {NoStop}%
\bibitem [{\citenamefont {Aharonovich}\ \emph {et~al.}(2016)\citenamefont
  {Aharonovich}, \citenamefont {Englund},\ and\ \citenamefont
  {Toth}}]{aharonovich_solid-state_2016}%
  \BibitemOpen
  \bibfield  {author} {\bibinfo {author} {\bibfnamefont {I.}~\bibnamefont
  {Aharonovich}}, \bibinfo {author} {\bibfnamefont {D.}~\bibnamefont
  {Englund}}, \ and\ \bibinfo {author} {\bibfnamefont {M.}~\bibnamefont
  {Toth}},\ }\href {\doibase 10.1038/nphoton.2016.186} {\bibfield  {journal}
  {\bibinfo  {journal} {Nature Photon}\ }\textbf {\bibinfo {volume} {10}},\
  \bibinfo {pages} {631} (\bibinfo {year} {2016})}\BibitemShut {NoStop}%
\bibitem [{\citenamefont {Atat{\"u}re}\ \emph {et~al.}(2018)\citenamefont
  {Atat{\"u}re}, \citenamefont {Englund}, \citenamefont {Vamivakas},
  \citenamefont {Lee},\ and\ \citenamefont
  {Wrachtrup}}]{atature_material_2018}%
  \BibitemOpen
  \bibfield  {author} {\bibinfo {author} {\bibfnamefont {M.}~\bibnamefont
  {Atat{\"u}re}}, \bibinfo {author} {\bibfnamefont {D.}~\bibnamefont
  {Englund}}, \bibinfo {author} {\bibfnamefont {N.}~\bibnamefont {Vamivakas}},
  \bibinfo {author} {\bibfnamefont {S.-Y.}\ \bibnamefont {Lee}}, \ and\
  \bibinfo {author} {\bibfnamefont {J.}~\bibnamefont {Wrachtrup}},\ }\href
  {\doibase 10.1038/s41578-018-0008-9} {\bibfield  {journal} {\bibinfo
  {journal} {Nat Rev Mater}\ }\textbf {\bibinfo {volume} {3}},\ \bibinfo
  {pages} {38} (\bibinfo {year} {2018})}\BibitemShut {NoStop}%
\bibitem [{\citenamefont {Mak}\ \emph {et~al.}(2013)\citenamefont {Mak},
  \citenamefont {He}, \citenamefont {Lee}, \citenamefont {Lee}, \citenamefont
  {Hone}, \citenamefont {Heinz},\ and\ \citenamefont
  {Shan}}]{mak_tightly_2013}%
  \BibitemOpen
  \bibfield  {author} {\bibinfo {author} {\bibfnamefont {K.~F.}\ \bibnamefont
  {Mak}}, \bibinfo {author} {\bibfnamefont {K.}~\bibnamefont {He}}, \bibinfo
  {author} {\bibfnamefont {C.}~\bibnamefont {Lee}}, \bibinfo {author}
  {\bibfnamefont {G.~H.}\ \bibnamefont {Lee}}, \bibinfo {author} {\bibfnamefont
  {J.}~\bibnamefont {Hone}}, \bibinfo {author} {\bibfnamefont {T.~F.}\
  \bibnamefont {Heinz}}, \ and\ \bibinfo {author} {\bibfnamefont
  {J.}~\bibnamefont {Shan}},\ }\href {\doibase 10.1038/nmat3505} {\bibfield
  {journal} {\bibinfo  {journal} {Nature Mater}\ }\textbf {\bibinfo {volume}
  {12}},\ \bibinfo {pages} {207} (\bibinfo {year} {2013})}\BibitemShut
  {NoStop}%
\bibitem [{\citenamefont {Barbone}\ \emph {et~al.}(2018)\citenamefont
  {Barbone}, \citenamefont {Montblanch}, \citenamefont {Kara}, \citenamefont
  {Palacios-Berraquero}, \citenamefont {Cadore}, \citenamefont {De~Fazio},
  \citenamefont {Pingault}, \citenamefont {Mostaani}, \citenamefont {Li},
  \citenamefont {Chen}, \citenamefont {Watanabe}, \citenamefont {Taniguchi},
  \citenamefont {Tongay}, \citenamefont {Wang}, \citenamefont {Ferrari},\ and\
  \citenamefont {Atat{\"u}re}}]{barbone_charge-tuneable_2018}%
  \BibitemOpen
  \bibfield  {author} {\bibinfo {author} {\bibfnamefont {M.}~\bibnamefont
  {Barbone}}, \bibinfo {author} {\bibfnamefont {A.~R.-P.}\ \bibnamefont
  {Montblanch}}, \bibinfo {author} {\bibfnamefont {D.~M.}\ \bibnamefont
  {Kara}}, \bibinfo {author} {\bibfnamefont {C.}~\bibnamefont
  {Palacios-Berraquero}}, \bibinfo {author} {\bibfnamefont {A.~R.}\
  \bibnamefont {Cadore}}, \bibinfo {author} {\bibfnamefont {D.}~\bibnamefont
  {De~Fazio}}, \bibinfo {author} {\bibfnamefont {B.}~\bibnamefont {Pingault}},
  \bibinfo {author} {\bibfnamefont {E.}~\bibnamefont {Mostaani}}, \bibinfo
  {author} {\bibfnamefont {H.}~\bibnamefont {Li}}, \bibinfo {author}
  {\bibfnamefont {B.}~\bibnamefont {Chen}}, \bibinfo {author} {\bibfnamefont
  {K.}~\bibnamefont {Watanabe}}, \bibinfo {author} {\bibfnamefont
  {T.}~\bibnamefont {Taniguchi}}, \bibinfo {author} {\bibfnamefont
  {S.}~\bibnamefont {Tongay}}, \bibinfo {author} {\bibfnamefont
  {G.}~\bibnamefont {Wang}}, \bibinfo {author} {\bibfnamefont {A.~C.}\
  \bibnamefont {Ferrari}}, \ and\ \bibinfo {author} {\bibfnamefont
  {M.}~\bibnamefont {Atat{\"u}re}},\ }\href {\doibase
  10.1038/s41467-018-05632-4} {\bibfield  {journal} {\bibinfo  {journal} {Nat
  Commun}\ }\textbf {\bibinfo {volume} {9}},\ \bibinfo {pages} {3721} (\bibinfo
  {year} {2018})}\BibitemShut {NoStop}%
\bibitem [{\citenamefont {Li}\ \emph {et~al.}(2018)\citenamefont {Li},
  \citenamefont {Wang}, \citenamefont {Lu}, \citenamefont {Jin}, \citenamefont
  {Chen}, \citenamefont {Meng}, \citenamefont {Lian}, \citenamefont
  {Taniguchi}, \citenamefont {Watanabe}, \citenamefont {Zhang}, \citenamefont
  {Smirnov},\ and\ \citenamefont {Shi}}]{li_revealing_2018}%
  \BibitemOpen
  \bibfield  {author} {\bibinfo {author} {\bibfnamefont {Z.}~\bibnamefont
  {Li}}, \bibinfo {author} {\bibfnamefont {T.}~\bibnamefont {Wang}}, \bibinfo
  {author} {\bibfnamefont {Z.}~\bibnamefont {Lu}}, \bibinfo {author}
  {\bibfnamefont {C.}~\bibnamefont {Jin}}, \bibinfo {author} {\bibfnamefont
  {Y.}~\bibnamefont {Chen}}, \bibinfo {author} {\bibfnamefont {Y.}~\bibnamefont
  {Meng}}, \bibinfo {author} {\bibfnamefont {Z.}~\bibnamefont {Lian}}, \bibinfo
  {author} {\bibfnamefont {T.}~\bibnamefont {Taniguchi}}, \bibinfo {author}
  {\bibfnamefont {K.}~\bibnamefont {Watanabe}}, \bibinfo {author}
  {\bibfnamefont {S.}~\bibnamefont {Zhang}}, \bibinfo {author} {\bibfnamefont
  {D.}~\bibnamefont {Smirnov}}, \ and\ \bibinfo {author} {\bibfnamefont
  {S.-F.}\ \bibnamefont {Shi}},\ }\href {\doibase 10.1038/s41467-018-05863-5}
  {\bibfield  {journal} {\bibinfo  {journal} {Nat Commun}\ }\textbf {\bibinfo
  {volume} {9}},\ \bibinfo {pages} {3719} (\bibinfo {year} {2018})}\BibitemShut
  {NoStop}%
\bibitem [{\citenamefont {Ye}\ \emph {et~al.}(2018)\citenamefont {Ye},
  \citenamefont {Waldecker}, \citenamefont {Ma}, \citenamefont {Rhodes},
  \citenamefont {Antony}, \citenamefont {Kim}, \citenamefont {Zhang},
  \citenamefont {Deng}, \citenamefont {Jiang}, \citenamefont {Lu},
  \citenamefont {Smirnov}, \citenamefont {Watanabe}, \citenamefont {Taniguchi},
  \citenamefont {Hone},\ and\ \citenamefont {Heinz}}]{ye_efficient_2018}%
  \BibitemOpen
  \bibfield  {author} {\bibinfo {author} {\bibfnamefont {Z.}~\bibnamefont
  {Ye}}, \bibinfo {author} {\bibfnamefont {L.}~\bibnamefont {Waldecker}},
  \bibinfo {author} {\bibfnamefont {E.~Y.}\ \bibnamefont {Ma}}, \bibinfo
  {author} {\bibfnamefont {D.}~\bibnamefont {Rhodes}}, \bibinfo {author}
  {\bibfnamefont {A.}~\bibnamefont {Antony}}, \bibinfo {author} {\bibfnamefont
  {B.}~\bibnamefont {Kim}}, \bibinfo {author} {\bibfnamefont {X.-X.}\
  \bibnamefont {Zhang}}, \bibinfo {author} {\bibfnamefont {M.}~\bibnamefont
  {Deng}}, \bibinfo {author} {\bibfnamefont {Y.}~\bibnamefont {Jiang}},
  \bibinfo {author} {\bibfnamefont {Z.}~\bibnamefont {Lu}}, \bibinfo {author}
  {\bibfnamefont {D.}~\bibnamefont {Smirnov}}, \bibinfo {author} {\bibfnamefont
  {K.}~\bibnamefont {Watanabe}}, \bibinfo {author} {\bibfnamefont
  {T.}~\bibnamefont {Taniguchi}}, \bibinfo {author} {\bibfnamefont
  {J.}~\bibnamefont {Hone}}, \ and\ \bibinfo {author} {\bibfnamefont {T.~F.}\
  \bibnamefont {Heinz}},\ }\href {\doibase 10.1038/s41467-018-05917-8}
  {\bibfield  {journal} {\bibinfo  {journal} {Nat Commun}\ }\textbf {\bibinfo
  {volume} {9}},\ \bibinfo {pages} {3718} (\bibinfo {year} {2018})}\BibitemShut
  {NoStop}%
\bibitem [{\citenamefont {Chen}\ \emph
  {et~al.}(2018{\natexlab{a}})\citenamefont {Chen}, \citenamefont {Goldstein},
  \citenamefont {Taniguchi}, \citenamefont {Watanabe},\ and\ \citenamefont
  {Yan}}]{chen_coulomb-bound_2018}%
  \BibitemOpen
  \bibfield  {author} {\bibinfo {author} {\bibfnamefont {S.-Y.}\ \bibnamefont
  {Chen}}, \bibinfo {author} {\bibfnamefont {T.}~\bibnamefont {Goldstein}},
  \bibinfo {author} {\bibfnamefont {T.}~\bibnamefont {Taniguchi}}, \bibinfo
  {author} {\bibfnamefont {K.}~\bibnamefont {Watanabe}}, \ and\ \bibinfo
  {author} {\bibfnamefont {J.}~\bibnamefont {Yan}},\ }\href {\doibase
  10.1038/s41467-018-05558-x} {\bibfield  {journal} {\bibinfo  {journal} {Nat
  Commun}\ }\textbf {\bibinfo {volume} {9}},\ \bibinfo {pages} {3717} (\bibinfo
  {year} {2018}{\natexlab{a}})}\BibitemShut {NoStop}%
\bibitem [{\citenamefont {Hao}\ \emph {et~al.}(2017)\citenamefont {Hao},
  \citenamefont {Specht}, \citenamefont {Nagler}, \citenamefont {Xu},
  \citenamefont {Tran}, \citenamefont {Singh}, \citenamefont {Dass},
  \citenamefont {Sch{\"u}ller}, \citenamefont {Korn}, \citenamefont {Richter},
  \citenamefont {Knorr}, \citenamefont {Li},\ and\ \citenamefont
  {Moody}}]{hao_neutral_2017}%
  \BibitemOpen
  \bibfield  {author} {\bibinfo {author} {\bibfnamefont {K.}~\bibnamefont
  {Hao}}, \bibinfo {author} {\bibfnamefont {J.~F.}\ \bibnamefont {Specht}},
  \bibinfo {author} {\bibfnamefont {P.}~\bibnamefont {Nagler}}, \bibinfo
  {author} {\bibfnamefont {L.}~\bibnamefont {Xu}}, \bibinfo {author}
  {\bibfnamefont {K.}~\bibnamefont {Tran}}, \bibinfo {author} {\bibfnamefont
  {A.}~\bibnamefont {Singh}}, \bibinfo {author} {\bibfnamefont {C.~K.}\
  \bibnamefont {Dass}}, \bibinfo {author} {\bibfnamefont {C.}~\bibnamefont
  {Sch{\"u}ller}}, \bibinfo {author} {\bibfnamefont {T.}~\bibnamefont {Korn}},
  \bibinfo {author} {\bibfnamefont {M.}~\bibnamefont {Richter}}, \bibinfo
  {author} {\bibfnamefont {A.}~\bibnamefont {Knorr}}, \bibinfo {author}
  {\bibfnamefont {X.}~\bibnamefont {Li}}, \ and\ \bibinfo {author}
  {\bibfnamefont {G.}~\bibnamefont {Moody}},\ }\href {\doibase
  10.1038/ncomms15552} {\bibfield  {journal} {\bibinfo  {journal} {Nat Commun}\
  }\textbf {\bibinfo {volume} {8}},\ \bibinfo {pages} {15552} (\bibinfo {year}
  {2017})}\BibitemShut {NoStop}%
\bibitem [{\citenamefont {Kremser}\ \emph {et~al.}(2020)\citenamefont
  {Kremser}, \citenamefont {Brotons-Gisbert}, \citenamefont {Kn{\"o}rzer},
  \citenamefont {G{\"u}ckelhorn}, \citenamefont {Meyer}, \citenamefont
  {Barbone}, \citenamefont {Stier}, \citenamefont {Gerardot}, \citenamefont
  {M{\"u}ller},\ and\ \citenamefont {Finley}}]{kremser_discrete_2020}%
  \BibitemOpen
  \bibfield  {author} {\bibinfo {author} {\bibfnamefont {M.}~\bibnamefont
  {Kremser}}, \bibinfo {author} {\bibfnamefont {M.}~\bibnamefont
  {Brotons-Gisbert}}, \bibinfo {author} {\bibfnamefont {J.}~\bibnamefont
  {Kn{\"o}rzer}}, \bibinfo {author} {\bibfnamefont {J.}~\bibnamefont
  {G{\"u}ckelhorn}}, \bibinfo {author} {\bibfnamefont {M.}~\bibnamefont
  {Meyer}}, \bibinfo {author} {\bibfnamefont {M.}~\bibnamefont {Barbone}},
  \bibinfo {author} {\bibfnamefont {A.~V.}\ \bibnamefont {Stier}}, \bibinfo
  {author} {\bibfnamefont {B.~D.}\ \bibnamefont {Gerardot}}, \bibinfo {author}
  {\bibfnamefont {K.}~\bibnamefont {M{\"u}ller}}, \ and\ \bibinfo {author}
  {\bibfnamefont {J.~J.}\ \bibnamefont {Finley}},\ }\href {\doibase
  10.1038/s41699-020-0141-3} {\bibfield  {journal} {\bibinfo  {journal} {npj 2D
  Mater Appl}\ }\textbf {\bibinfo {volume} {4}},\ \bibinfo {pages} {8}
  (\bibinfo {year} {2020})}\BibitemShut {NoStop}%
\bibitem [{\citenamefont {Klein}\ \emph
  {et~al.}(2021{\natexlab{b}})\citenamefont {Klein}, \citenamefont
  {H{\"o}tger}, \citenamefont {Florian}, \citenamefont {Steinhoff},
  \citenamefont {Delhomme}, \citenamefont {Taniguchi}, \citenamefont
  {Watanabe}, \citenamefont {Jahnke}, \citenamefont {Holleitner}, \citenamefont
  {Potemski}, \citenamefont {Faugeras}, \citenamefont {Finley},\ and\
  \citenamefont {Stier}}]{klein_controlling_2021}%
  \BibitemOpen
  \bibfield  {author} {\bibinfo {author} {\bibfnamefont {J.}~\bibnamefont
  {Klein}}, \bibinfo {author} {\bibfnamefont {A.}~\bibnamefont {H{\"o}tger}},
  \bibinfo {author} {\bibfnamefont {M.}~\bibnamefont {Florian}}, \bibinfo
  {author} {\bibfnamefont {A.}~\bibnamefont {Steinhoff}}, \bibinfo {author}
  {\bibfnamefont {A.}~\bibnamefont {Delhomme}}, \bibinfo {author}
  {\bibfnamefont {T.}~\bibnamefont {Taniguchi}}, \bibinfo {author}
  {\bibfnamefont {K.}~\bibnamefont {Watanabe}}, \bibinfo {author}
  {\bibfnamefont {F.}~\bibnamefont {Jahnke}}, \bibinfo {author} {\bibfnamefont
  {A.~W.}\ \bibnamefont {Holleitner}}, \bibinfo {author} {\bibfnamefont
  {M.}~\bibnamefont {Potemski}}, \bibinfo {author} {\bibfnamefont
  {C.}~\bibnamefont {Faugeras}}, \bibinfo {author} {\bibfnamefont {J.~J.}\
  \bibnamefont {Finley}}, \ and\ \bibinfo {author} {\bibfnamefont {A.~V.}\
  \bibnamefont {Stier}},\ }\href {\doibase 10.1103/PhysRevResearch.3.L022009}
  {\bibfield  {journal} {\bibinfo  {journal} {Phys. Rev. Research}\ }\textbf
  {\bibinfo {volume} {3}},\ \bibinfo {pages} {L022009} (\bibinfo {year}
  {2021}{\natexlab{b}})}\BibitemShut {NoStop}%
\bibitem [{\citenamefont {Costanzo}\ \emph {et~al.}(2016)\citenamefont
  {Costanzo}, \citenamefont {Jo}, \citenamefont {Berger},\ and\ \citenamefont
  {Morpurgo}}]{costanzo_gate-induced_2016}%
  \BibitemOpen
  \bibfield  {author} {\bibinfo {author} {\bibfnamefont {D.}~\bibnamefont
  {Costanzo}}, \bibinfo {author} {\bibfnamefont {S.}~\bibnamefont {Jo}},
  \bibinfo {author} {\bibfnamefont {H.}~\bibnamefont {Berger}}, \ and\ \bibinfo
  {author} {\bibfnamefont {A.~F.}\ \bibnamefont {Morpurgo}},\ }\href {\doibase
  10.1038/nnano.2015.314} {\bibfield  {journal} {\bibinfo  {journal} {Nature
  Nanotech}\ }\textbf {\bibinfo {volume} {11}},\ \bibinfo {pages} {339}
  (\bibinfo {year} {2016})}\BibitemShut {NoStop}%
\bibitem [{\citenamefont {Ye}\ \emph {et~al.}(2012)\citenamefont {Ye},
  \citenamefont {Zhang}, \citenamefont {Akashi}, \citenamefont {Bahramy},
  \citenamefont {Arita},\ and\ \citenamefont
  {Iwasa}}]{ye_superconducting_2012}%
  \BibitemOpen
  \bibfield  {author} {\bibinfo {author} {\bibfnamefont {J.~T.}\ \bibnamefont
  {Ye}}, \bibinfo {author} {\bibfnamefont {Y.~J.}\ \bibnamefont {Zhang}},
  \bibinfo {author} {\bibfnamefont {R.}~\bibnamefont {Akashi}}, \bibinfo
  {author} {\bibfnamefont {M.~S.}\ \bibnamefont {Bahramy}}, \bibinfo {author}
  {\bibfnamefont {R.}~\bibnamefont {Arita}}, \ and\ \bibinfo {author}
  {\bibfnamefont {Y.}~\bibnamefont {Iwasa}},\ }\href {\doibase
  10.1126/science.1228006} {\bibfield  {journal} {\bibinfo  {journal}
  {Science}\ }\textbf {\bibinfo {volume} {338}},\ \bibinfo {pages} {1193}
  (\bibinfo {year} {2012})}\BibitemShut {NoStop}%
\bibitem [{\citenamefont {Geim}\ and\ \citenamefont
  {Grigorieva}(2013)}]{geim_van_2013}%
  \BibitemOpen
  \bibfield  {author} {\bibinfo {author} {\bibfnamefont {A.~K.}\ \bibnamefont
  {Geim}}\ and\ \bibinfo {author} {\bibfnamefont {I.~V.}\ \bibnamefont
  {Grigorieva}},\ }\href {\doibase 10.1038/nature12385} {\bibfield  {journal}
  {\bibinfo  {journal} {Nature}\ }\textbf {\bibinfo {volume} {499}},\ \bibinfo
  {pages} {419} (\bibinfo {year} {2013})}\BibitemShut {NoStop}%
\bibitem [{\citenamefont {Rivera}\ \emph {et~al.}(2015)\citenamefont {Rivera},
  \citenamefont {Schaibley}, \citenamefont {Jones}, \citenamefont {Ross},
  \citenamefont {Wu}, \citenamefont {Aivazian}, \citenamefont {Klement},
  \citenamefont {Seyler}, \citenamefont {Clark}, \citenamefont {Ghimire},
  \citenamefont {Yan}, \citenamefont {Mandrus}, \citenamefont {Yao},\ and\
  \citenamefont {Xu}}]{rivera_observation_2015}%
  \BibitemOpen
  \bibfield  {author} {\bibinfo {author} {\bibfnamefont {P.}~\bibnamefont
  {Rivera}}, \bibinfo {author} {\bibfnamefont {J.~R.}\ \bibnamefont
  {Schaibley}}, \bibinfo {author} {\bibfnamefont {A.~M.}\ \bibnamefont
  {Jones}}, \bibinfo {author} {\bibfnamefont {J.~S.}\ \bibnamefont {Ross}},
  \bibinfo {author} {\bibfnamefont {S.}~\bibnamefont {Wu}}, \bibinfo {author}
  {\bibfnamefont {G.}~\bibnamefont {Aivazian}}, \bibinfo {author}
  {\bibfnamefont {P.}~\bibnamefont {Klement}}, \bibinfo {author} {\bibfnamefont
  {K.}~\bibnamefont {Seyler}}, \bibinfo {author} {\bibfnamefont
  {G.}~\bibnamefont {Clark}}, \bibinfo {author} {\bibfnamefont {N.~J.}\
  \bibnamefont {Ghimire}}, \bibinfo {author} {\bibfnamefont {J.}~\bibnamefont
  {Yan}}, \bibinfo {author} {\bibfnamefont {D.~G.}\ \bibnamefont {Mandrus}},
  \bibinfo {author} {\bibfnamefont {W.}~\bibnamefont {Yao}}, \ and\ \bibinfo
  {author} {\bibfnamefont {X.}~\bibnamefont {Xu}},\ }\href {\doibase
  10.1038/ncomms7242} {\bibfield  {journal} {\bibinfo  {journal} {Nat Commun}\
  }\textbf {\bibinfo {volume} {6}},\ \bibinfo {pages} {6242} (\bibinfo {year}
  {2015})}\BibitemShut {NoStop}%
\bibitem [{\citenamefont {Seyler}\ \emph {et~al.}(2019)\citenamefont {Seyler},
  \citenamefont {Rivera}, \citenamefont {Yu}, \citenamefont {Wilson},
  \citenamefont {Ray}, \citenamefont {Mandrus}, \citenamefont {Yan},
  \citenamefont {Yao},\ and\ \citenamefont {Xu}}]{seyler_signatures_2019}%
  \BibitemOpen
  \bibfield  {author} {\bibinfo {author} {\bibfnamefont {K.~L.}\ \bibnamefont
  {Seyler}}, \bibinfo {author} {\bibfnamefont {P.}~\bibnamefont {Rivera}},
  \bibinfo {author} {\bibfnamefont {H.}~\bibnamefont {Yu}}, \bibinfo {author}
  {\bibfnamefont {N.~P.}\ \bibnamefont {Wilson}}, \bibinfo {author}
  {\bibfnamefont {E.~L.}\ \bibnamefont {Ray}}, \bibinfo {author} {\bibfnamefont
  {D.~G.}\ \bibnamefont {Mandrus}}, \bibinfo {author} {\bibfnamefont
  {J.}~\bibnamefont {Yan}}, \bibinfo {author} {\bibfnamefont {W.}~\bibnamefont
  {Yao}}, \ and\ \bibinfo {author} {\bibfnamefont {X.}~\bibnamefont {Xu}},\
  }\href {\doibase 10.1038/s41586-019-0957-1} {\bibfield  {journal} {\bibinfo
  {journal} {Nature}\ }\textbf {\bibinfo {volume} {567}},\ \bibinfo {pages}
  {66} (\bibinfo {year} {2019})}\BibitemShut {NoStop}%
\bibitem [{\citenamefont {Tonndorf}\ \emph {et~al.}(2017)\citenamefont
  {Tonndorf}, \citenamefont {Del Pozo-Zamudio}, \citenamefont {Gruhler},
  \citenamefont {Kern}, \citenamefont {Schmidt}, \citenamefont {Dmitriev},
  \citenamefont {Bakhtinov}, \citenamefont {Tartakovskii}, \citenamefont
  {Pernice}, \citenamefont {Michaelis~de Vasconcellos},\ and\ \citenamefont
  {Bratschitsch}}]{tonndorf_-chip_2017}%
  \BibitemOpen
  \bibfield  {author} {\bibinfo {author} {\bibfnamefont {P.}~\bibnamefont
  {Tonndorf}}, \bibinfo {author} {\bibfnamefont {O.}~\bibnamefont {Del
  Pozo-Zamudio}}, \bibinfo {author} {\bibfnamefont {N.}~\bibnamefont
  {Gruhler}}, \bibinfo {author} {\bibfnamefont {J.}~\bibnamefont {Kern}},
  \bibinfo {author} {\bibfnamefont {R.}~\bibnamefont {Schmidt}}, \bibinfo
  {author} {\bibfnamefont {A.~I.}\ \bibnamefont {Dmitriev}}, \bibinfo {author}
  {\bibfnamefont {A.~P.}\ \bibnamefont {Bakhtinov}}, \bibinfo {author}
  {\bibfnamefont {A.~I.}\ \bibnamefont {Tartakovskii}}, \bibinfo {author}
  {\bibfnamefont {W.}~\bibnamefont {Pernice}}, \bibinfo {author} {\bibfnamefont
  {S.}~\bibnamefont {Michaelis~de Vasconcellos}}, \ and\ \bibinfo {author}
  {\bibfnamefont {R.}~\bibnamefont {Bratschitsch}},\ }\href {\doibase
  10.1021/acs.nanolett.7b02092} {\bibfield  {journal} {\bibinfo  {journal}
  {Nano Lett.}\ }\textbf {\bibinfo {volume} {17}},\ \bibinfo {pages} {5446}
  (\bibinfo {year} {2017})}\BibitemShut {NoStop}%
\bibitem [{\citenamefont {Youngblood}\ and\ \citenamefont
  {Li}(2016)}]{youngblood_integration_2016}%
  \BibitemOpen
  \bibfield  {author} {\bibinfo {author} {\bibfnamefont {N.}~\bibnamefont
  {Youngblood}}\ and\ \bibinfo {author} {\bibfnamefont {M.}~\bibnamefont
  {Li}},\ }\href {\doibase 10.1515/nanoph-2016-0155} {\bibfield  {journal}
  {\bibinfo  {journal} {Nanophotonics}\ }\textbf {\bibinfo {volume} {6}},\
  \bibinfo {pages} {1205} (\bibinfo {year} {2016})}\BibitemShut {NoStop}%
\bibitem [{\citenamefont {Sriram}\ \emph {et~al.}(2020)\citenamefont {Sriram},
  \citenamefont {Manikandan}, \citenamefont {Chuang},\ and\ \citenamefont
  {Chueh}}]{sriram_hybridizing_2020}%
  \BibitemOpen
  \bibfield  {author} {\bibinfo {author} {\bibfnamefont {P.}~\bibnamefont
  {Sriram}}, \bibinfo {author} {\bibfnamefont {A.}~\bibnamefont {Manikandan}},
  \bibinfo {author} {\bibfnamefont {F.-C.}\ \bibnamefont {Chuang}}, \ and\
  \bibinfo {author} {\bibfnamefont {Y.-L.}\ \bibnamefont {Chueh}},\ }\href
  {\doibase 10.1002/smll.201904271} {\bibfield  {journal} {\bibinfo  {journal}
  {Small}\ }\textbf {\bibinfo {volume} {16}},\ \bibinfo {pages} {1904271}
  (\bibinfo {year} {2020})}\BibitemShut {NoStop}%
\bibitem [{\citenamefont {Blauth}\ \emph {et~al.}(2017)\citenamefont {Blauth},
  \citenamefont {Harms}, \citenamefont {Prechtl}, \citenamefont {Finley},\ and\
  \citenamefont {Kaniber}}]{blauth_enhanced_2017}%
  \BibitemOpen
  \bibfield  {author} {\bibinfo {author} {\bibfnamefont {M.}~\bibnamefont
  {Blauth}}, \bibinfo {author} {\bibfnamefont {J.}~\bibnamefont {Harms}},
  \bibinfo {author} {\bibfnamefont {M.}~\bibnamefont {Prechtl}}, \bibinfo
  {author} {\bibfnamefont {J.~J.}\ \bibnamefont {Finley}}, \ and\ \bibinfo
  {author} {\bibfnamefont {M.}~\bibnamefont {Kaniber}},\ }\href {\doibase
  10.1088/2053-1583/aa52b0} {\bibfield  {journal} {\bibinfo  {journal} {2D
  Mater.}\ }\textbf {\bibinfo {volume} {4}},\ \bibinfo {pages} {021011}
  (\bibinfo {year} {2017})}\BibitemShut {NoStop}%
\bibitem [{\citenamefont {Blauth}\ \emph {et~al.}(2018)\citenamefont {Blauth},
  \citenamefont {J{\"u}rgensen}, \citenamefont {Vest}, \citenamefont {Hartwig},
  \citenamefont {Prechtl}, \citenamefont {Cerne}, \citenamefont {Finley},\ and\
  \citenamefont {Kaniber}}]{blauth_coupling_2018}%
  \BibitemOpen
  \bibfield  {author} {\bibinfo {author} {\bibfnamefont {M.}~\bibnamefont
  {Blauth}}, \bibinfo {author} {\bibfnamefont {M.}~\bibnamefont
  {J{\"u}rgensen}}, \bibinfo {author} {\bibfnamefont {G.}~\bibnamefont {Vest}},
  \bibinfo {author} {\bibfnamefont {O.}~\bibnamefont {Hartwig}}, \bibinfo
  {author} {\bibfnamefont {M.}~\bibnamefont {Prechtl}}, \bibinfo {author}
  {\bibfnamefont {J.}~\bibnamefont {Cerne}}, \bibinfo {author} {\bibfnamefont
  {J.~J.}\ \bibnamefont {Finley}}, \ and\ \bibinfo {author} {\bibfnamefont
  {M.}~\bibnamefont {Kaniber}},\ }\href {\doibase 10.1021/acs.nanolett.8b02687}
  {\bibfield  {journal} {\bibinfo  {journal} {Nano Lett.}\ }\textbf {\bibinfo
  {volume} {18}},\ \bibinfo {pages} {6812} (\bibinfo {year}
  {2018})}\BibitemShut {NoStop}%
\bibitem [{\citenamefont {Stier}\ \emph
  {et~al.}(2016{\natexlab{a}})\citenamefont {Stier}, \citenamefont {McCreary},
  \citenamefont {Jonker}, \citenamefont {Kono},\ and\ \citenamefont
  {Crooker}}]{stier_exciton_2016}%
  \BibitemOpen
  \bibfield  {author} {\bibinfo {author} {\bibfnamefont {A.~V.}\ \bibnamefont
  {Stier}}, \bibinfo {author} {\bibfnamefont {K.~M.}\ \bibnamefont {McCreary}},
  \bibinfo {author} {\bibfnamefont {B.~T.}\ \bibnamefont {Jonker}}, \bibinfo
  {author} {\bibfnamefont {J.}~\bibnamefont {Kono}}, \ and\ \bibinfo {author}
  {\bibfnamefont {S.~A.}\ \bibnamefont {Crooker}},\ }\href {\doibase
  10.1038/ncomms10643} {\bibfield  {journal} {\bibinfo  {journal} {Nat Commun}\
  }\textbf {\bibinfo {volume} {7}},\ \bibinfo {pages} {10643} (\bibinfo {year}
  {2016}{\natexlab{a}})}\BibitemShut {NoStop}%
\bibitem [{\citenamefont {Stier}\ \emph {et~al.}(2018)\citenamefont {Stier},
  \citenamefont {Wilson}, \citenamefont {Velizhanin}, \citenamefont {Kono},
  \citenamefont {Xu},\ and\ \citenamefont
  {Crooker}}]{stier_magnetooptics_2018}%
  \BibitemOpen
  \bibfield  {author} {\bibinfo {author} {\bibfnamefont {A.~V.}\ \bibnamefont
  {Stier}}, \bibinfo {author} {\bibfnamefont {N.~P.}\ \bibnamefont {Wilson}},
  \bibinfo {author} {\bibfnamefont {K.~A.}\ \bibnamefont {Velizhanin}},
  \bibinfo {author} {\bibfnamefont {J.}~\bibnamefont {Kono}}, \bibinfo {author}
  {\bibfnamefont {X.}~\bibnamefont {Xu}}, \ and\ \bibinfo {author}
  {\bibfnamefont {S.~A.}\ \bibnamefont {Crooker}},\ }\href {\doibase
  10.1103/PhysRevLett.120.057405} {\bibfield  {journal} {\bibinfo  {journal}
  {Phys. Rev. Lett.}\ }\textbf {\bibinfo {volume} {120}},\ \bibinfo {pages}
  {057405} (\bibinfo {year} {2018})}\BibitemShut {NoStop}%
\bibitem [{\citenamefont {Goryca}\ \emph {et~al.}(2019)\citenamefont {Goryca},
  \citenamefont {Li}, \citenamefont {Stier}, \citenamefont {Taniguchi},
  \citenamefont {Watanabe}, \citenamefont {Courtade}, \citenamefont {Shree},
  \citenamefont {Robert}, \citenamefont {Urbaszek}, \citenamefont {Marie},\
  and\ \citenamefont {Crooker}}]{goryca_revealing_2019}%
  \BibitemOpen
  \bibfield  {author} {\bibinfo {author} {\bibfnamefont {M.}~\bibnamefont
  {Goryca}}, \bibinfo {author} {\bibfnamefont {J.}~\bibnamefont {Li}}, \bibinfo
  {author} {\bibfnamefont {A.~V.}\ \bibnamefont {Stier}}, \bibinfo {author}
  {\bibfnamefont {T.}~\bibnamefont {Taniguchi}}, \bibinfo {author}
  {\bibfnamefont {K.}~\bibnamefont {Watanabe}}, \bibinfo {author}
  {\bibfnamefont {E.}~\bibnamefont {Courtade}}, \bibinfo {author}
  {\bibfnamefont {S.}~\bibnamefont {Shree}}, \bibinfo {author} {\bibfnamefont
  {C.}~\bibnamefont {Robert}}, \bibinfo {author} {\bibfnamefont
  {B.}~\bibnamefont {Urbaszek}}, \bibinfo {author} {\bibfnamefont
  {X.}~\bibnamefont {Marie}}, \ and\ \bibinfo {author} {\bibfnamefont {S.~A.}\
  \bibnamefont {Crooker}},\ }\href {\doibase 10.1038/s41467-019-12180-y}
  {\bibfield  {journal} {\bibinfo  {journal} {Nat Commun}\ }\textbf {\bibinfo
  {volume} {10}},\ \bibinfo {pages} {4172} (\bibinfo {year}
  {2019})}\BibitemShut {NoStop}%
\bibitem [{\citenamefont {Baranov}\ \emph {et~al.}(2018)\citenamefont
  {Baranov}, \citenamefont {Wers{\"a}ll}, \citenamefont {Cuadra}, \citenamefont
  {Antosiewicz},\ and\ \citenamefont {Shegai}}]{baranov_novel_2018}%
  \BibitemOpen
  \bibfield  {author} {\bibinfo {author} {\bibfnamefont {D.~G.}\ \bibnamefont
  {Baranov}}, \bibinfo {author} {\bibfnamefont {M.}~\bibnamefont
  {Wers{\"a}ll}}, \bibinfo {author} {\bibfnamefont {J.}~\bibnamefont {Cuadra}},
  \bibinfo {author} {\bibfnamefont {T.~J.}\ \bibnamefont {Antosiewicz}}, \ and\
  \bibinfo {author} {\bibfnamefont {T.}~\bibnamefont {Shegai}},\ }\href
  {\doibase 10.1021/acsphotonics.7b00674} {\bibfield  {journal} {\bibinfo
  {journal} {ACS Photonics}\ }\textbf {\bibinfo {volume} {5}},\ \bibinfo
  {pages} {24} (\bibinfo {year} {2018})}\BibitemShut {NoStop}%
\bibitem [{\citenamefont {Pelton}\ \emph {et~al.}(2019)\citenamefont {Pelton},
  \citenamefont {Storm},\ and\ \citenamefont {Leng}}]{pelton_strong_2019}%
  \BibitemOpen
  \bibfield  {author} {\bibinfo {author} {\bibfnamefont {M.}~\bibnamefont
  {Pelton}}, \bibinfo {author} {\bibfnamefont {S.~D.}\ \bibnamefont {Storm}}, \
  and\ \bibinfo {author} {\bibfnamefont {H.}~\bibnamefont {Leng}},\ }\href
  {\doibase 10.1039/C9NR05044B} {\bibfield  {journal} {\bibinfo  {journal}
  {Nanoscale}\ }\textbf {\bibinfo {volume} {11}},\ \bibinfo {pages} {14540}
  (\bibinfo {year} {2019})}\BibitemShut {NoStop}%
\bibitem [{\citenamefont {Akselrod}\ \emph {et~al.}(2014)\citenamefont
  {Akselrod}, \citenamefont {Argyropoulos}, \citenamefont {Hoang},
  \citenamefont {Cirac{\`i}}, \citenamefont {Fang}, \citenamefont {Huang},
  \citenamefont {Smith},\ and\ \citenamefont
  {Mikkelsen}}]{akselrod_probing_2014}%
  \BibitemOpen
  \bibfield  {author} {\bibinfo {author} {\bibfnamefont {G.~M.}\ \bibnamefont
  {Akselrod}}, \bibinfo {author} {\bibfnamefont {C.}~\bibnamefont
  {Argyropoulos}}, \bibinfo {author} {\bibfnamefont {T.~B.}\ \bibnamefont
  {Hoang}}, \bibinfo {author} {\bibfnamefont {C.}~\bibnamefont {Cirac{\`i}}},
  \bibinfo {author} {\bibfnamefont {C.}~\bibnamefont {Fang}}, \bibinfo {author}
  {\bibfnamefont {J.}~\bibnamefont {Huang}}, \bibinfo {author} {\bibfnamefont
  {D.~R.}\ \bibnamefont {Smith}}, \ and\ \bibinfo {author} {\bibfnamefont
  {M.~H.}\ \bibnamefont {Mikkelsen}},\ }\href {\doibase
  10.1038/nphoton.2014.228} {\bibfield  {journal} {\bibinfo  {journal} {Nature
  Photon}\ }\textbf {\bibinfo {volume} {8}},\ \bibinfo {pages} {835} (\bibinfo
  {year} {2014})}\BibitemShut {NoStop}%
\bibitem [{\citenamefont {Chikkaraddy}\ \emph {et~al.}(2016)\citenamefont
  {Chikkaraddy}, \citenamefont {de~Nijs}, \citenamefont {Benz}, \citenamefont
  {Barrow}, \citenamefont {Scherman}, \citenamefont {Rosta}, \citenamefont
  {Demetriadou}, \citenamefont {Fox}, \citenamefont {Hess},\ and\ \citenamefont
  {Baumberg}}]{chikkaraddy_single-molecule_2016}%
  \BibitemOpen
  \bibfield  {author} {\bibinfo {author} {\bibfnamefont {R.}~\bibnamefont
  {Chikkaraddy}}, \bibinfo {author} {\bibfnamefont {B.}~\bibnamefont
  {de~Nijs}}, \bibinfo {author} {\bibfnamefont {F.}~\bibnamefont {Benz}},
  \bibinfo {author} {\bibfnamefont {S.~J.}\ \bibnamefont {Barrow}}, \bibinfo
  {author} {\bibfnamefont {O.~A.}\ \bibnamefont {Scherman}}, \bibinfo {author}
  {\bibfnamefont {E.}~\bibnamefont {Rosta}}, \bibinfo {author} {\bibfnamefont
  {A.}~\bibnamefont {Demetriadou}}, \bibinfo {author} {\bibfnamefont
  {P.}~\bibnamefont {Fox}}, \bibinfo {author} {\bibfnamefont {O.}~\bibnamefont
  {Hess}}, \ and\ \bibinfo {author} {\bibfnamefont {J.~J.}\ \bibnamefont
  {Baumberg}},\ }\href {\doibase 10.1038/nature17974} {\bibfield  {journal}
  {\bibinfo  {journal} {Nature}\ }\textbf {\bibinfo {volume} {535}},\ \bibinfo
  {pages} {127} (\bibinfo {year} {2016})}\BibitemShut {NoStop}%
\bibitem [{\citenamefont {Miroshnichenko}\ \emph {et~al.}(2010)\citenamefont
  {Miroshnichenko}, \citenamefont {Flach},\ and\ \citenamefont
  {Kivshar}}]{miroshnichenko_fano_2010}%
  \BibitemOpen
  \bibfield  {author} {\bibinfo {author} {\bibfnamefont {A.~E.}\ \bibnamefont
  {Miroshnichenko}}, \bibinfo {author} {\bibfnamefont {S.}~\bibnamefont
  {Flach}}, \ and\ \bibinfo {author} {\bibfnamefont {Y.~S.}\ \bibnamefont
  {Kivshar}},\ }\href {\doibase 10.1103/RevModPhys.82.2257} {\bibfield
  {journal} {\bibinfo  {journal} {Rev. Mod. Phys.}\ }\textbf {\bibinfo {volume}
  {82}},\ \bibinfo {pages} {2257} (\bibinfo {year} {2010})}\BibitemShut
  {NoStop}%
\bibitem [{\citenamefont {Lee}\ \emph {et~al.}(2015)\citenamefont {Lee},
  \citenamefont {Park}, \citenamefont {Han}, \citenamefont {Ee}, \citenamefont
  {Naylor}, \citenamefont {Liu}, \citenamefont {Johnson},\ and\ \citenamefont
  {Agarwal}}]{lee_fano_2015}%
  \BibitemOpen
  \bibfield  {author} {\bibinfo {author} {\bibfnamefont {B.}~\bibnamefont
  {Lee}}, \bibinfo {author} {\bibfnamefont {J.}~\bibnamefont {Park}}, \bibinfo
  {author} {\bibfnamefont {G.~H.}\ \bibnamefont {Han}}, \bibinfo {author}
  {\bibfnamefont {H.-S.}\ \bibnamefont {Ee}}, \bibinfo {author} {\bibfnamefont
  {C.~H.}\ \bibnamefont {Naylor}}, \bibinfo {author} {\bibfnamefont
  {W.}~\bibnamefont {Liu}}, \bibinfo {author} {\bibfnamefont {A.~C.}\
  \bibnamefont {Johnson}}, \ and\ \bibinfo {author} {\bibfnamefont
  {R.}~\bibnamefont {Agarwal}},\ }\href {\doibase 10.1021/acs.nanolett.5b01563}
  {\bibfield  {journal} {\bibinfo  {journal} {Nano Lett.}\ }\textbf {\bibinfo
  {volume} {15}},\ \bibinfo {pages} {3646} (\bibinfo {year}
  {2015})}\BibitemShut {NoStop}%
\bibitem [{\citenamefont {Abid}\ \emph {et~al.}(2017)\citenamefont {Abid},
  \citenamefont {Chen}, \citenamefont {Yuan}, \citenamefont {Bohloul},
  \citenamefont {Najmaei}, \citenamefont {Avendano}, \citenamefont
  {P{\'e}chou}, \citenamefont {Mlayah},\ and\ \citenamefont
  {Lou}}]{abid_temperature-dependent_2017}%
  \BibitemOpen
  \bibfield  {author} {\bibinfo {author} {\bibfnamefont {I.}~\bibnamefont
  {Abid}}, \bibinfo {author} {\bibfnamefont {W.}~\bibnamefont {Chen}}, \bibinfo
  {author} {\bibfnamefont {J.}~\bibnamefont {Yuan}}, \bibinfo {author}
  {\bibfnamefont {A.}~\bibnamefont {Bohloul}}, \bibinfo {author} {\bibfnamefont
  {S.}~\bibnamefont {Najmaei}}, \bibinfo {author} {\bibfnamefont
  {C.}~\bibnamefont {Avendano}}, \bibinfo {author} {\bibfnamefont
  {R.}~\bibnamefont {P{\'e}chou}}, \bibinfo {author} {\bibfnamefont
  {A.}~\bibnamefont {Mlayah}}, \ and\ \bibinfo {author} {\bibfnamefont
  {J.}~\bibnamefont {Lou}},\ }\href {\doibase 10.1021/acsphotonics.6b00957}
  {\bibfield  {journal} {\bibinfo  {journal} {ACS Photonics}\ }\textbf
  {\bibinfo {volume} {4}},\ \bibinfo {pages} {1653} (\bibinfo {year}
  {2017})}\BibitemShut {NoStop}%
\bibitem [{\citenamefont {Wang}\ \emph
  {et~al.}(2018{\natexlab{b}})\citenamefont {Wang}, \citenamefont {Krasnok},
  \citenamefont {Zhang}, \citenamefont {Scarabelli}, \citenamefont {Liu},
  \citenamefont {Wu}, \citenamefont {Liz-Marz{\'a}n}, \citenamefont {Terrones},
  \citenamefont {Al{\`u}},\ and\ \citenamefont {Zheng}}]{wang_tunable_2018}%
  \BibitemOpen
  \bibfield  {author} {\bibinfo {author} {\bibfnamefont {M.}~\bibnamefont
  {Wang}}, \bibinfo {author} {\bibfnamefont {A.}~\bibnamefont {Krasnok}},
  \bibinfo {author} {\bibfnamefont {T.}~\bibnamefont {Zhang}}, \bibinfo
  {author} {\bibfnamefont {L.}~\bibnamefont {Scarabelli}}, \bibinfo {author}
  {\bibfnamefont {H.}~\bibnamefont {Liu}}, \bibinfo {author} {\bibfnamefont
  {Z.}~\bibnamefont {Wu}}, \bibinfo {author} {\bibfnamefont {L.~M.}\
  \bibnamefont {Liz-Marz{\'a}n}}, \bibinfo {author} {\bibfnamefont
  {M.}~\bibnamefont {Terrones}}, \bibinfo {author} {\bibfnamefont
  {A.}~\bibnamefont {Al{\`u}}}, \ and\ \bibinfo {author} {\bibfnamefont
  {Y.}~\bibnamefont {Zheng}},\ }\href {\doibase 10.1002/adma.201705779}
  {\bibfield  {journal} {\bibinfo  {journal} {Adv. Mater.}\ }\textbf {\bibinfo
  {volume} {30}},\ \bibinfo {pages} {1705779} (\bibinfo {year}
  {2018}{\natexlab{b}})}\BibitemShut {NoStop}%
\bibitem [{\citenamefont {Sun}\ \emph {et~al.}(2018)\citenamefont {Sun},
  \citenamefont {Hu}, \citenamefont {Zheng}, \citenamefont {Zhang},
  \citenamefont {Deng}, \citenamefont {Zhang},\ and\ \citenamefont
  {Xu}}]{sun_light-emitting_2018}%
  \BibitemOpen
  \bibfield  {author} {\bibinfo {author} {\bibfnamefont {J.}~\bibnamefont
  {Sun}}, \bibinfo {author} {\bibfnamefont {H.}~\bibnamefont {Hu}}, \bibinfo
  {author} {\bibfnamefont {D.}~\bibnamefont {Zheng}}, \bibinfo {author}
  {\bibfnamefont {D.}~\bibnamefont {Zhang}}, \bibinfo {author} {\bibfnamefont
  {Q.}~\bibnamefont {Deng}}, \bibinfo {author} {\bibfnamefont {S.}~\bibnamefont
  {Zhang}}, \ and\ \bibinfo {author} {\bibfnamefont {H.}~\bibnamefont {Xu}},\
  }\href {\doibase 10.1021/acsnano.8b05880} {\bibfield  {journal} {\bibinfo
  {journal} {ACS Nano}\ }\textbf {\bibinfo {volume} {12}},\ \bibinfo {pages}
  {10393} (\bibinfo {year} {2018})}\BibitemShut {NoStop}%
\bibitem [{\citenamefont {Kleemann}\ \emph {et~al.}(2017)\citenamefont
  {Kleemann}, \citenamefont {Chikkaraddy}, \citenamefont {Alexeev},
  \citenamefont {Kos}, \citenamefont {Carnegie}, \citenamefont {Deacon},
  \citenamefont {de~Pury}, \citenamefont {Gro{\ss}e}, \citenamefont {de~Nijs},
  \citenamefont {Mertens}, \citenamefont {Tartakovskii},\ and\ \citenamefont
  {Baumberg}}]{kleemann_strong-coupling_2017}%
  \BibitemOpen
  \bibfield  {author} {\bibinfo {author} {\bibfnamefont {M.-E.}\ \bibnamefont
  {Kleemann}}, \bibinfo {author} {\bibfnamefont {R.}~\bibnamefont
  {Chikkaraddy}}, \bibinfo {author} {\bibfnamefont {E.~M.}\ \bibnamefont
  {Alexeev}}, \bibinfo {author} {\bibfnamefont {D.}~\bibnamefont {Kos}},
  \bibinfo {author} {\bibfnamefont {C.}~\bibnamefont {Carnegie}}, \bibinfo
  {author} {\bibfnamefont {W.}~\bibnamefont {Deacon}}, \bibinfo {author}
  {\bibfnamefont {A.~C.}\ \bibnamefont {de~Pury}}, \bibinfo {author}
  {\bibfnamefont {C.}~\bibnamefont {Gro{\ss}e}}, \bibinfo {author}
  {\bibfnamefont {B.}~\bibnamefont {de~Nijs}}, \bibinfo {author} {\bibfnamefont
  {J.}~\bibnamefont {Mertens}}, \bibinfo {author} {\bibfnamefont {A.~I.}\
  \bibnamefont {Tartakovskii}}, \ and\ \bibinfo {author} {\bibfnamefont
  {J.~J.}\ \bibnamefont {Baumberg}},\ }\href {\doibase
  10.1038/s41467-017-01398-3} {\bibfield  {journal} {\bibinfo  {journal} {Nat
  Commun}\ }\textbf {\bibinfo {volume} {8}},\ \bibinfo {pages} {1296} (\bibinfo
  {year} {2017})}\BibitemShut {NoStop}%
\bibitem [{\citenamefont {Wen}\ \emph {et~al.}(2017)\citenamefont {Wen},
  \citenamefont {Wang}, \citenamefont {Wang}, \citenamefont {Deng},
  \citenamefont {Zhuang}, \citenamefont {Zhang}, \citenamefont {Liu},
  \citenamefont {She}, \citenamefont {Chen}, \citenamefont {Chen},
  \citenamefont {Deng},\ and\ \citenamefont {Xu}}]{wen_room-temperature_2017}%
  \BibitemOpen
  \bibfield  {author} {\bibinfo {author} {\bibfnamefont {J.}~\bibnamefont
  {Wen}}, \bibinfo {author} {\bibfnamefont {H.}~\bibnamefont {Wang}}, \bibinfo
  {author} {\bibfnamefont {W.}~\bibnamefont {Wang}}, \bibinfo {author}
  {\bibfnamefont {Z.}~\bibnamefont {Deng}}, \bibinfo {author} {\bibfnamefont
  {C.}~\bibnamefont {Zhuang}}, \bibinfo {author} {\bibfnamefont
  {Y.}~\bibnamefont {Zhang}}, \bibinfo {author} {\bibfnamefont
  {F.}~\bibnamefont {Liu}}, \bibinfo {author} {\bibfnamefont {J.}~\bibnamefont
  {She}}, \bibinfo {author} {\bibfnamefont {J.}~\bibnamefont {Chen}}, \bibinfo
  {author} {\bibfnamefont {H.}~\bibnamefont {Chen}}, \bibinfo {author}
  {\bibfnamefont {S.}~\bibnamefont {Deng}}, \ and\ \bibinfo {author}
  {\bibfnamefont {N.}~\bibnamefont {Xu}},\ }\href {\doibase
  10.1021/acs.nanolett.7b01344} {\bibfield  {journal} {\bibinfo  {journal}
  {Nano Lett.}\ }\textbf {\bibinfo {volume} {17}},\ \bibinfo {pages} {4689}
  (\bibinfo {year} {2017})}\BibitemShut {NoStop}%
\bibitem [{\citenamefont {Zheng}\ \emph {et~al.}(2017)\citenamefont {Zheng},
  \citenamefont {Zhang}, \citenamefont {Deng}, \citenamefont {Kang},
  \citenamefont {Nordlander},\ and\ \citenamefont
  {Xu}}]{zheng_manipulating_2017}%
  \BibitemOpen
  \bibfield  {author} {\bibinfo {author} {\bibfnamefont {D.}~\bibnamefont
  {Zheng}}, \bibinfo {author} {\bibfnamefont {S.}~\bibnamefont {Zhang}},
  \bibinfo {author} {\bibfnamefont {Q.}~\bibnamefont {Deng}}, \bibinfo {author}
  {\bibfnamefont {M.}~\bibnamefont {Kang}}, \bibinfo {author} {\bibfnamefont
  {P.}~\bibnamefont {Nordlander}}, \ and\ \bibinfo {author} {\bibfnamefont
  {H.}~\bibnamefont {Xu}},\ }\href {\doibase 10.1021/acs.nanolett.7b01176}
  {\bibfield  {journal} {\bibinfo  {journal} {Nano Lett.}\ }\textbf {\bibinfo
  {volume} {17}},\ \bibinfo {pages} {3809} (\bibinfo {year}
  {2017})}\BibitemShut {NoStop}%
\bibitem [{\citenamefont {Geisler}\ \emph {et~al.}(2019)\citenamefont
  {Geisler}, \citenamefont {Cui}, \citenamefont {Wang}, \citenamefont
  {Rindzevicius}, \citenamefont {Gammelgaard}, \citenamefont {Jessen},
  \citenamefont {Gon{\c c}alves}, \citenamefont {Todisco}, \citenamefont
  {B{\o}ggild}, \citenamefont {Boisen}, \citenamefont {Wubs}, \citenamefont
  {Mortensen}, \citenamefont {Xiao},\ and\ \citenamefont
  {Stenger}}]{geisler_single-crystalline_2019}%
  \BibitemOpen
  \bibfield  {author} {\bibinfo {author} {\bibfnamefont {M.}~\bibnamefont
  {Geisler}}, \bibinfo {author} {\bibfnamefont {X.}~\bibnamefont {Cui}},
  \bibinfo {author} {\bibfnamefont {J.}~\bibnamefont {Wang}}, \bibinfo {author}
  {\bibfnamefont {T.}~\bibnamefont {Rindzevicius}}, \bibinfo {author}
  {\bibfnamefont {L.}~\bibnamefont {Gammelgaard}}, \bibinfo {author}
  {\bibfnamefont {B.~S.}\ \bibnamefont {Jessen}}, \bibinfo {author}
  {\bibfnamefont {P.~A.~D.}\ \bibnamefont {Gon{\c c}alves}}, \bibinfo {author}
  {\bibfnamefont {F.}~\bibnamefont {Todisco}}, \bibinfo {author} {\bibfnamefont
  {P.}~\bibnamefont {B{\o}ggild}}, \bibinfo {author} {\bibfnamefont
  {A.}~\bibnamefont {Boisen}}, \bibinfo {author} {\bibfnamefont
  {M.}~\bibnamefont {Wubs}}, \bibinfo {author} {\bibfnamefont {N.~A.}\
  \bibnamefont {Mortensen}}, \bibinfo {author} {\bibfnamefont {S.}~\bibnamefont
  {Xiao}}, \ and\ \bibinfo {author} {\bibfnamefont {N.}~\bibnamefont
  {Stenger}},\ }\href {\doibase 10.1021/acsphotonics.8b01766} {\bibfield
  {journal} {\bibinfo  {journal} {ACS Photonics}\ }\textbf {\bibinfo {volume}
  {6}},\ \bibinfo {pages} {994} (\bibinfo {year} {2019})}\BibitemShut {NoStop}%
\bibitem [{\citenamefont {Qin}\ \emph {et~al.}(2020)\citenamefont {Qin},
  \citenamefont {Chen}, \citenamefont {Zhang}, \citenamefont {Zhang},
  \citenamefont {Blaikie}, \citenamefont {Ding},\ and\ \citenamefont
  {Qiu}}]{qin_revealing_2020}%
  \BibitemOpen
  \bibfield  {author} {\bibinfo {author} {\bibfnamefont {J.}~\bibnamefont
  {Qin}}, \bibinfo {author} {\bibfnamefont {Y.-H.}\ \bibnamefont {Chen}},
  \bibinfo {author} {\bibfnamefont {Z.}~\bibnamefont {Zhang}}, \bibinfo
  {author} {\bibfnamefont {Y.}~\bibnamefont {Zhang}}, \bibinfo {author}
  {\bibfnamefont {R.~J.}\ \bibnamefont {Blaikie}}, \bibinfo {author}
  {\bibfnamefont {B.}~\bibnamefont {Ding}}, \ and\ \bibinfo {author}
  {\bibfnamefont {M.}~\bibnamefont {Qiu}},\ }\href {\doibase
  10.1103/PhysRevLett.124.063902} {\bibfield  {journal} {\bibinfo  {journal}
  {Phys. Rev. Lett.}\ }\textbf {\bibinfo {volume} {124}},\ \bibinfo {pages}
  {063902} (\bibinfo {year} {2020})}\BibitemShut {NoStop}%
\bibitem [{\citenamefont {Wang}\ \emph {et~al.}(2016)\citenamefont {Wang},
  \citenamefont {Li}, \citenamefont {Chervy}, \citenamefont {Shalabney},
  \citenamefont {Azzini}, \citenamefont {Orgiu}, \citenamefont {Hutchison},
  \citenamefont {Genet}, \citenamefont {Samor{\`i}},\ and\ \citenamefont
  {Ebbesen}}]{wang_coherent_2016}%
  \BibitemOpen
  \bibfield  {author} {\bibinfo {author} {\bibfnamefont {S.}~\bibnamefont
  {Wang}}, \bibinfo {author} {\bibfnamefont {S.}~\bibnamefont {Li}}, \bibinfo
  {author} {\bibfnamefont {T.}~\bibnamefont {Chervy}}, \bibinfo {author}
  {\bibfnamefont {A.}~\bibnamefont {Shalabney}}, \bibinfo {author}
  {\bibfnamefont {S.}~\bibnamefont {Azzini}}, \bibinfo {author} {\bibfnamefont
  {E.}~\bibnamefont {Orgiu}}, \bibinfo {author} {\bibfnamefont {J.~A.}\
  \bibnamefont {Hutchison}}, \bibinfo {author} {\bibfnamefont {C.}~\bibnamefont
  {Genet}}, \bibinfo {author} {\bibfnamefont {P.}~\bibnamefont {Samor{\`i}}}, \
  and\ \bibinfo {author} {\bibfnamefont {T.~W.}\ \bibnamefont {Ebbesen}},\
  }\href {\doibase 10.1021/acs.nanolett.6b01475} {\bibfield  {journal}
  {\bibinfo  {journal} {Nano Lett.}\ }\textbf {\bibinfo {volume} {16}},\
  \bibinfo {pages} {4368} (\bibinfo {year} {2016})}\BibitemShut {NoStop}%
\bibitem [{\citenamefont {Liu}\ \emph {et~al.}(2016)\citenamefont {Liu},
  \citenamefont {Lee}, \citenamefont {Naylor}, \citenamefont {Ee},
  \citenamefont {Park}, \citenamefont {Johnson},\ and\ \citenamefont
  {Agarwal}}]{liu_strong_2016}%
  \BibitemOpen
  \bibfield  {author} {\bibinfo {author} {\bibfnamefont {W.}~\bibnamefont
  {Liu}}, \bibinfo {author} {\bibfnamefont {B.}~\bibnamefont {Lee}}, \bibinfo
  {author} {\bibfnamefont {C.~H.}\ \bibnamefont {Naylor}}, \bibinfo {author}
  {\bibfnamefont {H.-S.}\ \bibnamefont {Ee}}, \bibinfo {author} {\bibfnamefont
  {J.}~\bibnamefont {Park}}, \bibinfo {author} {\bibfnamefont {A.~T.~C.}\
  \bibnamefont {Johnson}}, \ and\ \bibinfo {author} {\bibfnamefont
  {R.}~\bibnamefont {Agarwal}},\ }\href {\doibase 10.1021/acs.nanolett.5b04588}
  {\bibfield  {journal} {\bibinfo  {journal} {Nano Lett.}\ }\textbf {\bibinfo
  {volume} {16}},\ \bibinfo {pages} {1262} (\bibinfo {year}
  {2016})}\BibitemShut {NoStop}%
\bibitem [{\citenamefont {Sortino}\ \emph {et~al.}(2019)\citenamefont
  {Sortino}, \citenamefont {Zotev}, \citenamefont {Mignuzzi}, \citenamefont
  {Cambiasso}, \citenamefont {Schmidt}, \citenamefont {Genco}, \citenamefont
  {A{\ss}mann}, \citenamefont {Bayer}, \citenamefont {Maier}, \citenamefont
  {Sapienza},\ and\ \citenamefont {Tartakovskii}}]{sortino_enhanced_2019}%
  \BibitemOpen
  \bibfield  {author} {\bibinfo {author} {\bibfnamefont {L.}~\bibnamefont
  {Sortino}}, \bibinfo {author} {\bibfnamefont {P.~G.}\ \bibnamefont {Zotev}},
  \bibinfo {author} {\bibfnamefont {S.}~\bibnamefont {Mignuzzi}}, \bibinfo
  {author} {\bibfnamefont {J.}~\bibnamefont {Cambiasso}}, \bibinfo {author}
  {\bibfnamefont {D.}~\bibnamefont {Schmidt}}, \bibinfo {author} {\bibfnamefont
  {A.}~\bibnamefont {Genco}}, \bibinfo {author} {\bibfnamefont
  {M.}~\bibnamefont {A{\ss}mann}}, \bibinfo {author} {\bibfnamefont
  {M.}~\bibnamefont {Bayer}}, \bibinfo {author} {\bibfnamefont {S.~A.}\
  \bibnamefont {Maier}}, \bibinfo {author} {\bibfnamefont {R.}~\bibnamefont
  {Sapienza}}, \ and\ \bibinfo {author} {\bibfnamefont {A.~I.}\ \bibnamefont
  {Tartakovskii}},\ }\href {\doibase 10.1038/s41467-019-12963-3} {\bibfield
  {journal} {\bibinfo  {journal} {Nat Commun}\ }\textbf {\bibinfo {volume}
  {10}},\ \bibinfo {pages} {5119} (\bibinfo {year} {2019})}\BibitemShut
  {NoStop}%
\bibitem [{\citenamefont {Yan}\ and\ \citenamefont
  {Wei}(2020)}]{yan_strong_2020}%
  \BibitemOpen
  \bibfield  {author} {\bibinfo {author} {\bibfnamefont {X.}~\bibnamefont
  {Yan}}\ and\ \bibinfo {author} {\bibfnamefont {H.}~\bibnamefont {Wei}},\
  }\href {\doibase 10.1039/D0NR01056A} {\bibfield  {journal} {\bibinfo
  {journal} {Nanoscale}\ }\textbf {\bibinfo {volume} {12}},\ \bibinfo {pages}
  {9708} (\bibinfo {year} {2020})}\BibitemShut {NoStop}%
\bibitem [{\citenamefont {Luo}\ \emph {et~al.}(2018)\citenamefont {Luo},
  \citenamefont {Shepard}, \citenamefont {Ardelean}, \citenamefont {Rhodes},
  \citenamefont {Kim}, \citenamefont {Barmak}, \citenamefont {Hone},\ and\
  \citenamefont {Strauf}}]{luo_deterministic_2018}%
  \BibitemOpen
  \bibfield  {author} {\bibinfo {author} {\bibfnamefont {Y.}~\bibnamefont
  {Luo}}, \bibinfo {author} {\bibfnamefont {G.~D.}\ \bibnamefont {Shepard}},
  \bibinfo {author} {\bibfnamefont {J.~V.}\ \bibnamefont {Ardelean}}, \bibinfo
  {author} {\bibfnamefont {D.~A.}\ \bibnamefont {Rhodes}}, \bibinfo {author}
  {\bibfnamefont {B.}~\bibnamefont {Kim}}, \bibinfo {author} {\bibfnamefont
  {K.}~\bibnamefont {Barmak}}, \bibinfo {author} {\bibfnamefont {J.~C.}\
  \bibnamefont {Hone}}, \ and\ \bibinfo {author} {\bibfnamefont
  {S.}~\bibnamefont {Strauf}},\ }\href {\doibase 10.1038/s41565-018-0275-z}
  {\bibfield  {journal} {\bibinfo  {journal} {Nature Nanotech}\ }\textbf
  {\bibinfo {volume} {13}},\ \bibinfo {pages} {1137} (\bibinfo {year}
  {2018})}\BibitemShut {NoStop}%
\bibitem [{\citenamefont {Cai}\ \emph {et~al.}(2018)\citenamefont {Cai},
  \citenamefont {Kim}, \citenamefont {Yang}, \citenamefont {Dutta},
  \citenamefont {Aghaeimeibodi},\ and\ \citenamefont
  {Waks}}]{cai_radiative_2018}%
  \BibitemOpen
  \bibfield  {author} {\bibinfo {author} {\bibfnamefont {T.}~\bibnamefont
  {Cai}}, \bibinfo {author} {\bibfnamefont {J.-H.}\ \bibnamefont {Kim}},
  \bibinfo {author} {\bibfnamefont {Z.}~\bibnamefont {Yang}}, \bibinfo {author}
  {\bibfnamefont {S.}~\bibnamefont {Dutta}}, \bibinfo {author} {\bibfnamefont
  {S.}~\bibnamefont {Aghaeimeibodi}}, \ and\ \bibinfo {author} {\bibfnamefont
  {E.}~\bibnamefont {Waks}},\ }\href {\doibase 10.1021/acsphotonics.8b00580}
  {\bibfield  {journal} {\bibinfo  {journal} {ACS Photonics}\ }\textbf
  {\bibinfo {volume} {5}},\ \bibinfo {pages} {3466} (\bibinfo {year}
  {2018})}\BibitemShut {NoStop}%
\bibitem [{\citenamefont {Temnov}\ and\ \citenamefont
  {Woggon}(2005)}]{temnov_superradiance_2005}%
  \BibitemOpen
  \bibfield  {author} {\bibinfo {author} {\bibfnamefont {V.~V.}\ \bibnamefont
  {Temnov}}\ and\ \bibinfo {author} {\bibfnamefont {U.}~\bibnamefont
  {Woggon}},\ }\href {\doibase 10.1103/PhysRevLett.95.243602} {\bibfield
  {journal} {\bibinfo  {journal} {Phys. Rev. Lett.}\ }\textbf {\bibinfo
  {volume} {95}},\ \bibinfo {pages} {243602} (\bibinfo {year}
  {2005})}\BibitemShut {NoStop}%
\bibitem [{\citenamefont {Stobbe}\ \emph {et~al.}(2012)\citenamefont {Stobbe},
  \citenamefont {Kristensen}, \citenamefont {Mortensen}, \citenamefont {Hvam},
  \citenamefont {M{\o}rk},\ and\ \citenamefont
  {Lodahl}}]{stobbe_spontaneous_2012}%
  \BibitemOpen
  \bibfield  {author} {\bibinfo {author} {\bibfnamefont {S.}~\bibnamefont
  {Stobbe}}, \bibinfo {author} {\bibfnamefont {P.~T.}\ \bibnamefont
  {Kristensen}}, \bibinfo {author} {\bibfnamefont {J.~E.}\ \bibnamefont
  {Mortensen}}, \bibinfo {author} {\bibfnamefont {J.~M.}\ \bibnamefont {Hvam}},
  \bibinfo {author} {\bibfnamefont {J.}~\bibnamefont {M{\o}rk}}, \ and\
  \bibinfo {author} {\bibfnamefont {P.}~\bibnamefont {Lodahl}},\ }\href
  {\doibase 10.1103/PhysRevB.86.085304} {\bibfield  {journal} {\bibinfo
  {journal} {Phys. Rev. B}\ }\textbf {\bibinfo {volume} {86}},\ \bibinfo
  {pages} {085304} (\bibinfo {year} {2012})}\BibitemShut {NoStop}%
\bibitem [{\citenamefont {Novotny}\ and\ \citenamefont {van
  Hulst}(2011)}]{novotny_antennas_2011}%
  \BibitemOpen
  \bibfield  {author} {\bibinfo {author} {\bibfnamefont {L.}~\bibnamefont
  {Novotny}}\ and\ \bibinfo {author} {\bibfnamefont {N.}~\bibnamefont {van
  Hulst}},\ }\href {\doibase 10.1038/nphoton.2010.237} {\bibfield  {journal}
  {\bibinfo  {journal} {Nature Photon}\ }\textbf {\bibinfo {volume} {5}},\
  \bibinfo {pages} {83} (\bibinfo {year} {2011})}\BibitemShut {NoStop}%
\bibitem [{\citenamefont {Kinkhabwala}\ \emph {et~al.}(2009)\citenamefont
  {Kinkhabwala}, \citenamefont {Yu}, \citenamefont {Fan}, \citenamefont
  {Avlasevich}, \citenamefont {M{\"u}llen},\ and\ \citenamefont
  {Moerner}}]{kinkhabwala_large_2009}%
  \BibitemOpen
  \bibfield  {author} {\bibinfo {author} {\bibfnamefont {A.}~\bibnamefont
  {Kinkhabwala}}, \bibinfo {author} {\bibfnamefont {Z.}~\bibnamefont {Yu}},
  \bibinfo {author} {\bibfnamefont {S.}~\bibnamefont {Fan}}, \bibinfo {author}
  {\bibfnamefont {Y.}~\bibnamefont {Avlasevich}}, \bibinfo {author}
  {\bibfnamefont {K.}~\bibnamefont {M{\"u}llen}}, \ and\ \bibinfo {author}
  {\bibfnamefont {W.~E.}\ \bibnamefont {Moerner}},\ }\href {\doibase
  10.1038/nphoton.2009.187} {\bibfield  {journal} {\bibinfo  {journal} {Nature
  Photon}\ }\textbf {\bibinfo {volume} {3}},\ \bibinfo {pages} {654} (\bibinfo
  {year} {2009})}\BibitemShut {NoStop}%
\bibitem [{\citenamefont {Su}\ \emph {et~al.}(2003)\citenamefont {Su},
  \citenamefont {Wei}, \citenamefont {Zhang}, \citenamefont {Mock},
  \citenamefont {Smith},\ and\ \citenamefont
  {Schultz}}]{su_interparticle_2003}%
  \BibitemOpen
  \bibfield  {author} {\bibinfo {author} {\bibfnamefont {K.-H.}\ \bibnamefont
  {Su}}, \bibinfo {author} {\bibfnamefont {Q.-H.}\ \bibnamefont {Wei}},
  \bibinfo {author} {\bibfnamefont {X.}~\bibnamefont {Zhang}}, \bibinfo
  {author} {\bibfnamefont {J.~J.}\ \bibnamefont {Mock}}, \bibinfo {author}
  {\bibfnamefont {D.~R.}\ \bibnamefont {Smith}}, \ and\ \bibinfo {author}
  {\bibfnamefont {S.}~\bibnamefont {Schultz}},\ }\href {\doibase
  10.1021/nl034197f} {\bibfield  {journal} {\bibinfo  {journal} {Nano Lett.}\
  }\textbf {\bibinfo {volume} {3}},\ \bibinfo {pages} {1087} (\bibinfo {year}
  {2003})}\BibitemShut {NoStop}%
\bibitem [{\citenamefont {Liao}\ and\ \citenamefont
  {Wokaun}(1982)}]{liao_lightning_1982}%
  \BibitemOpen
  \bibfield  {author} {\bibinfo {author} {\bibfnamefont {P.~F.}\ \bibnamefont
  {Liao}}\ and\ \bibinfo {author} {\bibfnamefont {A.}~\bibnamefont {Wokaun}},\
  }\href {\doibase 10.1063/1.442690} {\bibfield  {journal} {\bibinfo  {journal}
  {The Journal of Chemical Physics}\ }\textbf {\bibinfo {volume} {76}},\
  \bibinfo {pages} {751} (\bibinfo {year} {1982})}\BibitemShut {NoStop}%
\bibitem [{\citenamefont {Schraml}\ \emph {et~al.}(2014)\citenamefont
  {Schraml}, \citenamefont {Spiegl}, \citenamefont {Kammerlocher},
  \citenamefont {Bracher}, \citenamefont {Bartl}, \citenamefont {Campbell},
  \citenamefont {Finley},\ and\ \citenamefont
  {Kaniber}}]{schraml_optical_2014}%
  \BibitemOpen
  \bibfield  {author} {\bibinfo {author} {\bibfnamefont {K.}~\bibnamefont
  {Schraml}}, \bibinfo {author} {\bibfnamefont {M.}~\bibnamefont {Spiegl}},
  \bibinfo {author} {\bibfnamefont {M.}~\bibnamefont {Kammerlocher}}, \bibinfo
  {author} {\bibfnamefont {G.}~\bibnamefont {Bracher}}, \bibinfo {author}
  {\bibfnamefont {J.}~\bibnamefont {Bartl}}, \bibinfo {author} {\bibfnamefont
  {T.}~\bibnamefont {Campbell}}, \bibinfo {author} {\bibfnamefont {J.~J.}\
  \bibnamefont {Finley}}, \ and\ \bibinfo {author} {\bibfnamefont
  {M.}~\bibnamefont {Kaniber}},\ }\href {\doibase 10.1103/PhysRevB.90.035435}
  {\bibfield  {journal} {\bibinfo  {journal} {Phys. Rev. B}\ }\textbf {\bibinfo
  {volume} {90}},\ \bibinfo {pages} {035435} (\bibinfo {year}
  {2014})}\BibitemShut {NoStop}%
\bibitem [{\citenamefont {Castellanos-Gomez}\ \emph {et~al.}(2014)\citenamefont
  {Castellanos-Gomez}, \citenamefont {Buscema}, \citenamefont {Molenaar},
  \citenamefont {Singh}, \citenamefont {Janssen}, \citenamefont {van~der
  Zant},\ and\ \citenamefont {Steele}}]{castellanos-gomez_deterministic_2014}%
  \BibitemOpen
  \bibfield  {author} {\bibinfo {author} {\bibfnamefont {A.}~\bibnamefont
  {Castellanos-Gomez}}, \bibinfo {author} {\bibfnamefont {M.}~\bibnamefont
  {Buscema}}, \bibinfo {author} {\bibfnamefont {R.}~\bibnamefont {Molenaar}},
  \bibinfo {author} {\bibfnamefont {V.}~\bibnamefont {Singh}}, \bibinfo
  {author} {\bibfnamefont {L.}~\bibnamefont {Janssen}}, \bibinfo {author}
  {\bibfnamefont {H.~S.~J.}\ \bibnamefont {van~der Zant}}, \ and\ \bibinfo
  {author} {\bibfnamefont {G.~A.}\ \bibnamefont {Steele}},\ }\href {\doibase
  10.1088/2053-1583/1/1/011002} {\bibfield  {journal} {\bibinfo  {journal} {2D
  Mater.}\ }\textbf {\bibinfo {volume} {1}},\ \bibinfo {pages} {011002}
  (\bibinfo {year} {2014})}\BibitemShut {NoStop}%
\bibitem [{\citenamefont {Li}\ \emph {et~al.}(2014)\citenamefont {Li},
  \citenamefont {Chernikov}, \citenamefont {Zhang}, \citenamefont {Rigosi},
  \citenamefont {Hill}, \citenamefont {van~der Zande}, \citenamefont {Chenet},
  \citenamefont {Shih}, \citenamefont {Hone},\ and\ \citenamefont
  {Heinz}}]{li_measurement_2014}%
  \BibitemOpen
  \bibfield  {author} {\bibinfo {author} {\bibfnamefont {Y.}~\bibnamefont
  {Li}}, \bibinfo {author} {\bibfnamefont {A.}~\bibnamefont {Chernikov}},
  \bibinfo {author} {\bibfnamefont {X.}~\bibnamefont {Zhang}}, \bibinfo
  {author} {\bibfnamefont {A.}~\bibnamefont {Rigosi}}, \bibinfo {author}
  {\bibfnamefont {H.~M.}\ \bibnamefont {Hill}}, \bibinfo {author}
  {\bibfnamefont {A.~M.}\ \bibnamefont {van~der Zande}}, \bibinfo {author}
  {\bibfnamefont {D.~A.}\ \bibnamefont {Chenet}}, \bibinfo {author}
  {\bibfnamefont {E.-M.}\ \bibnamefont {Shih}}, \bibinfo {author}
  {\bibfnamefont {J.}~\bibnamefont {Hone}}, \ and\ \bibinfo {author}
  {\bibfnamefont {T.~F.}\ \bibnamefont {Heinz}},\ }\href {\doibase
  10.1103/PhysRevB.90.205422} {\bibfield  {journal} {\bibinfo  {journal} {Phys.
  Rev. B}\ }\textbf {\bibinfo {volume} {90}},\ \bibinfo {pages} {205422}
  (\bibinfo {year} {2014})}\BibitemShut {NoStop}%
\bibitem [{\citenamefont {Wu}\ \emph {et~al.}(2010)\citenamefont {Wu},
  \citenamefont {Gray},\ and\ \citenamefont
  {Pelton}}]{wu_quantum-dot-induced_2010}%
  \BibitemOpen
  \bibfield  {author} {\bibinfo {author} {\bibfnamefont {X.}~\bibnamefont
  {Wu}}, \bibinfo {author} {\bibfnamefont {S.~K.}\ \bibnamefont {Gray}}, \ and\
  \bibinfo {author} {\bibfnamefont {M.}~\bibnamefont {Pelton}},\ }\href
  {\doibase 10.1364/OE.18.023633} {\bibfield  {journal} {\bibinfo  {journal}
  {Opt. Express}\ }\textbf {\bibinfo {volume} {18}},\ \bibinfo {pages} {23633}
  (\bibinfo {year} {2010})}\BibitemShut {NoStop}%
\bibitem [{\citenamefont {Limonov}\ \emph {et~al.}(2017)\citenamefont
  {Limonov}, \citenamefont {Rybin}, \citenamefont {Poddubny},\ and\
  \citenamefont {Kivshar}}]{limonov_fano_2017}%
  \BibitemOpen
  \bibfield  {author} {\bibinfo {author} {\bibfnamefont {M.~F.}\ \bibnamefont
  {Limonov}}, \bibinfo {author} {\bibfnamefont {M.~V.}\ \bibnamefont {Rybin}},
  \bibinfo {author} {\bibfnamefont {A.~N.}\ \bibnamefont {Poddubny}}, \ and\
  \bibinfo {author} {\bibfnamefont {Y.~S.}\ \bibnamefont {Kivshar}},\ }\href
  {\doibase 10.1038/nphoton.2017.142} {\bibfield  {journal} {\bibinfo
  {journal} {Nature Photon}\ }\textbf {\bibinfo {volume} {11}},\ \bibinfo
  {pages} {543} (\bibinfo {year} {2017})}\BibitemShut {NoStop}%
\bibitem [{\citenamefont {R{\"o}sner}\ \emph {et~al.}(2016)\citenamefont
  {R{\"o}sner}, \citenamefont {Steinke}, \citenamefont {Lorke}, \citenamefont
  {Gies}, \citenamefont {Jahnke},\ and\ \citenamefont
  {Wehling}}]{rosner_two-dimensional_2016}%
  \BibitemOpen
  \bibfield  {author} {\bibinfo {author} {\bibfnamefont {M.}~\bibnamefont
  {R{\"o}sner}}, \bibinfo {author} {\bibfnamefont {C.}~\bibnamefont {Steinke}},
  \bibinfo {author} {\bibfnamefont {M.}~\bibnamefont {Lorke}}, \bibinfo
  {author} {\bibfnamefont {C.}~\bibnamefont {Gies}}, \bibinfo {author}
  {\bibfnamefont {F.}~\bibnamefont {Jahnke}}, \ and\ \bibinfo {author}
  {\bibfnamefont {T.~O.}\ \bibnamefont {Wehling}},\ }\href {\doibase
  10.1021/acs.nanolett.5b05009} {\bibfield  {journal} {\bibinfo  {journal}
  {Nano Lett.}\ }\textbf {\bibinfo {volume} {16}},\ \bibinfo {pages} {2322}
  (\bibinfo {year} {2016})}\BibitemShut {NoStop}%
\bibitem [{\citenamefont {Stier}\ \emph
  {et~al.}(2016{\natexlab{b}})\citenamefont {Stier}, \citenamefont {Wilson},
  \citenamefont {Clark}, \citenamefont {Xu},\ and\ \citenamefont
  {Crooker}}]{stier_probing_2016}%
  \BibitemOpen
  \bibfield  {author} {\bibinfo {author} {\bibfnamefont {A.~V.}\ \bibnamefont
  {Stier}}, \bibinfo {author} {\bibfnamefont {N.~P.}\ \bibnamefont {Wilson}},
  \bibinfo {author} {\bibfnamefont {G.}~\bibnamefont {Clark}}, \bibinfo
  {author} {\bibfnamefont {X.}~\bibnamefont {Xu}}, \ and\ \bibinfo {author}
  {\bibfnamefont {S.~A.}\ \bibnamefont {Crooker}},\ }\href {\doibase
  10.1021/acs.nanolett.6b03276} {\bibfield  {journal} {\bibinfo  {journal}
  {Nano Lett.}\ }\textbf {\bibinfo {volume} {16}},\ \bibinfo {pages} {7054}
  (\bibinfo {year} {2016}{\natexlab{b}})}\BibitemShut {NoStop}%
\bibitem [{\citenamefont {Ugeda}\ \emph {et~al.}(2014)\citenamefont {Ugeda},
  \citenamefont {Bradley}, \citenamefont {Shi}, \citenamefont {da~Jornada},
  \citenamefont {Zhang}, \citenamefont {Qiu}, \citenamefont {Ruan},
  \citenamefont {Mo}, \citenamefont {Hussain}, \citenamefont {Shen},
  \citenamefont {Wang}, \citenamefont {Louie},\ and\ \citenamefont
  {Crommie}}]{ugeda_giant_2014}%
  \BibitemOpen
  \bibfield  {author} {\bibinfo {author} {\bibfnamefont {M.~M.}\ \bibnamefont
  {Ugeda}}, \bibinfo {author} {\bibfnamefont {A.~J.}\ \bibnamefont {Bradley}},
  \bibinfo {author} {\bibfnamefont {S.-F.}\ \bibnamefont {Shi}}, \bibinfo
  {author} {\bibfnamefont {F.~H.}\ \bibnamefont {da~Jornada}}, \bibinfo
  {author} {\bibfnamefont {Y.}~\bibnamefont {Zhang}}, \bibinfo {author}
  {\bibfnamefont {D.~Y.}\ \bibnamefont {Qiu}}, \bibinfo {author} {\bibfnamefont
  {W.}~\bibnamefont {Ruan}}, \bibinfo {author} {\bibfnamefont {S.-K.}\
  \bibnamefont {Mo}}, \bibinfo {author} {\bibfnamefont {Z.}~\bibnamefont
  {Hussain}}, \bibinfo {author} {\bibfnamefont {Z.-X.}\ \bibnamefont {Shen}},
  \bibinfo {author} {\bibfnamefont {F.}~\bibnamefont {Wang}}, \bibinfo {author}
  {\bibfnamefont {S.~G.}\ \bibnamefont {Louie}}, \ and\ \bibinfo {author}
  {\bibfnamefont {M.~F.}\ \bibnamefont {Crommie}},\ }\href {\doibase
  10.1038/nmat4061} {\bibfield  {journal} {\bibinfo  {journal} {Nature Mater}\
  }\textbf {\bibinfo {volume} {13}},\ \bibinfo {pages} {1091} (\bibinfo {year}
  {2014})}\BibitemShut {NoStop}%
\bibitem [{\citenamefont {Schmidt}\ \emph {et~al.}(2016)\citenamefont
  {Schmidt}, \citenamefont {Niehues}, \citenamefont {Schneider}, \citenamefont
  {Dr{\"u}ppel}, \citenamefont {Deilmann}, \citenamefont {Rohlfing},
  \citenamefont {de~Vasconcellos}, \citenamefont {Castellanos-Gomez},\ and\
  \citenamefont {Bratschitsch}}]{schmidt_reversible_2016}%
  \BibitemOpen
  \bibfield  {author} {\bibinfo {author} {\bibfnamefont {R.}~\bibnamefont
  {Schmidt}}, \bibinfo {author} {\bibfnamefont {I.}~\bibnamefont {Niehues}},
  \bibinfo {author} {\bibfnamefont {R.}~\bibnamefont {Schneider}}, \bibinfo
  {author} {\bibfnamefont {M.}~\bibnamefont {Dr{\"u}ppel}}, \bibinfo {author}
  {\bibfnamefont {T.}~\bibnamefont {Deilmann}}, \bibinfo {author}
  {\bibfnamefont {M.}~\bibnamefont {Rohlfing}}, \bibinfo {author}
  {\bibfnamefont {S.~M.}\ \bibnamefont {de~Vasconcellos}}, \bibinfo {author}
  {\bibfnamefont {A.}~\bibnamefont {Castellanos-Gomez}}, \ and\ \bibinfo
  {author} {\bibfnamefont {R.}~\bibnamefont {Bratschitsch}},\ }\href {\doibase
  10.1088/2053-1583/3/2/021011} {\bibfield  {journal} {\bibinfo  {journal} {2D
  Mater.}\ }\textbf {\bibinfo {volume} {3}},\ \bibinfo {pages} {021011}
  (\bibinfo {year} {2016})}\BibitemShut {NoStop}%
\bibitem [{\citenamefont {Branny}\ \emph {et~al.}(2016)\citenamefont {Branny},
  \citenamefont {Wang}, \citenamefont {Kumar}, \citenamefont {Robert},
  \citenamefont {Lassagne}, \citenamefont {Marie}, \citenamefont {Gerardot},\
  and\ \citenamefont {Urbaszek}}]{branny_discrete_2016}%
  \BibitemOpen
  \bibfield  {author} {\bibinfo {author} {\bibfnamefont {A.}~\bibnamefont
  {Branny}}, \bibinfo {author} {\bibfnamefont {G.}~\bibnamefont {Wang}},
  \bibinfo {author} {\bibfnamefont {S.}~\bibnamefont {Kumar}}, \bibinfo
  {author} {\bibfnamefont {C.}~\bibnamefont {Robert}}, \bibinfo {author}
  {\bibfnamefont {B.}~\bibnamefont {Lassagne}}, \bibinfo {author}
  {\bibfnamefont {X.}~\bibnamefont {Marie}}, \bibinfo {author} {\bibfnamefont
  {B.~D.}\ \bibnamefont {Gerardot}}, \ and\ \bibinfo {author} {\bibfnamefont
  {B.}~\bibnamefont {Urbaszek}},\ }\href {\doibase 10.1063/1.4945268}
  {\bibfield  {journal} {\bibinfo  {journal} {Appl. Phys. Lett.}\ }\textbf
  {\bibinfo {volume} {108}},\ \bibinfo {pages} {142101} (\bibinfo {year}
  {2016})}\BibitemShut {NoStop}%
\bibitem [{\citenamefont {Yu}\ \emph {et~al.}(2021)\citenamefont {Yu},
  \citenamefont {Deng}, \citenamefont {Zhang}, \citenamefont {Borghardt},
  \citenamefont {Kardynal}, \citenamefont {Vu{\v c}kovi{\'c}},\ and\
  \citenamefont {Heinz}}]{yu_site-controlled_2021}%
  \BibitemOpen
  \bibfield  {author} {\bibinfo {author} {\bibfnamefont {L.}~\bibnamefont
  {Yu}}, \bibinfo {author} {\bibfnamefont {M.}~\bibnamefont {Deng}}, \bibinfo
  {author} {\bibfnamefont {J.~L.}\ \bibnamefont {Zhang}}, \bibinfo {author}
  {\bibfnamefont {S.}~\bibnamefont {Borghardt}}, \bibinfo {author}
  {\bibfnamefont {B.}~\bibnamefont {Kardynal}}, \bibinfo {author}
  {\bibfnamefont {J.}~\bibnamefont {Vu{\v c}kovi{\'c}}}, \ and\ \bibinfo
  {author} {\bibfnamefont {T.~F.}\ \bibnamefont {Heinz}},\ }\href {\doibase
  10.1021/acs.nanolett.0c04282} {\bibfield  {journal} {\bibinfo  {journal}
  {Nano Lett.}\ }\textbf {\bibinfo {volume} {21}},\ \bibinfo {pages} {2376}
  (\bibinfo {year} {2021})}\BibitemShut {NoStop}%
\bibitem [{\citenamefont {Klein}\ \emph {et~al.}(2019)\citenamefont {Klein},
  \citenamefont {Lorke}, \citenamefont {Florian}, \citenamefont {Sigger},
  \citenamefont {Sigl}, \citenamefont {Rey}, \citenamefont {Wierzbowski},
  \citenamefont {Cerne}, \citenamefont {M{\"u}ller}, \citenamefont
  {Mitterreiter}, \citenamefont {Zimmermann}, \citenamefont {Taniguchi},
  \citenamefont {Watanabe}, \citenamefont {Wurstbauer}, \citenamefont
  {Kaniber}, \citenamefont {Knap}, \citenamefont {Schmidt}, \citenamefont
  {Finley},\ and\ \citenamefont {Holleitner}}]{klein_site-selectively_2019}%
  \BibitemOpen
  \bibfield  {author} {\bibinfo {author} {\bibfnamefont {J.}~\bibnamefont
  {Klein}}, \bibinfo {author} {\bibfnamefont {M.}~\bibnamefont {Lorke}},
  \bibinfo {author} {\bibfnamefont {M.}~\bibnamefont {Florian}}, \bibinfo
  {author} {\bibfnamefont {F.}~\bibnamefont {Sigger}}, \bibinfo {author}
  {\bibfnamefont {L.}~\bibnamefont {Sigl}}, \bibinfo {author} {\bibfnamefont
  {S.}~\bibnamefont {Rey}}, \bibinfo {author} {\bibfnamefont {J.}~\bibnamefont
  {Wierzbowski}}, \bibinfo {author} {\bibfnamefont {J.}~\bibnamefont {Cerne}},
  \bibinfo {author} {\bibfnamefont {K.}~\bibnamefont {M{\"u}ller}}, \bibinfo
  {author} {\bibfnamefont {E.}~\bibnamefont {Mitterreiter}}, \bibinfo {author}
  {\bibfnamefont {P.}~\bibnamefont {Zimmermann}}, \bibinfo {author}
  {\bibfnamefont {T.}~\bibnamefont {Taniguchi}}, \bibinfo {author}
  {\bibfnamefont {K.}~\bibnamefont {Watanabe}}, \bibinfo {author}
  {\bibfnamefont {U.}~\bibnamefont {Wurstbauer}}, \bibinfo {author}
  {\bibfnamefont {M.}~\bibnamefont {Kaniber}}, \bibinfo {author} {\bibfnamefont
  {M.}~\bibnamefont {Knap}}, \bibinfo {author} {\bibfnamefont {R.}~\bibnamefont
  {Schmidt}}, \bibinfo {author} {\bibfnamefont {J.~J.}\ \bibnamefont {Finley}},
  \ and\ \bibinfo {author} {\bibfnamefont {A.~W.}\ \bibnamefont {Holleitner}},\
  }\href {\doibase 10.1038/s41467-019-10632-z} {\bibfield  {journal} {\bibinfo
  {journal} {Nat Commun}\ }\textbf {\bibinfo {volume} {10}},\ \bibinfo {pages}
  {2755} (\bibinfo {year} {2019})}\BibitemShut {NoStop}%
\bibitem [{\citenamefont {Chen}\ \emph
  {et~al.}(2018{\natexlab{b}})\citenamefont {Chen}, \citenamefont {Razinskas},
  \citenamefont {Vieker}, \citenamefont {Gross}, \citenamefont {Wu},
  \citenamefont {Beyer}, \citenamefont {G{\"o}lzh{\"a}user},\ and\
  \citenamefont {Hecht}}]{chen_high-_2018}%
  \BibitemOpen
  \bibfield  {author} {\bibinfo {author} {\bibfnamefont {K.}~\bibnamefont
  {Chen}}, \bibinfo {author} {\bibfnamefont {G.}~\bibnamefont {Razinskas}},
  \bibinfo {author} {\bibfnamefont {H.}~\bibnamefont {Vieker}}, \bibinfo
  {author} {\bibfnamefont {H.}~\bibnamefont {Gross}}, \bibinfo {author}
  {\bibfnamefont {X.}~\bibnamefont {Wu}}, \bibinfo {author} {\bibfnamefont
  {A.}~\bibnamefont {Beyer}}, \bibinfo {author} {\bibfnamefont
  {A.}~\bibnamefont {G{\"o}lzh{\"a}user}}, \ and\ \bibinfo {author}
  {\bibfnamefont {B.}~\bibnamefont {Hecht}},\ }\href {\doibase
  10.1039/C8NR02160K} {\bibfield  {journal} {\bibinfo  {journal} {Nanoscale}\
  }\textbf {\bibinfo {volume} {10}},\ \bibinfo {pages} {17148} (\bibinfo {year}
  {2018}{\natexlab{b}})}\BibitemShut {NoStop}%
\bibitem [{\citenamefont {Kern}\ \emph {et~al.}(2015)\citenamefont {Kern},
  \citenamefont {Kullock}, \citenamefont {Prangsma}, \citenamefont {Emmerling},
  \citenamefont {Kamp},\ and\ \citenamefont {Hecht}}]{kern_electrically_2015}%
  \BibitemOpen
  \bibfield  {author} {\bibinfo {author} {\bibfnamefont {J.}~\bibnamefont
  {Kern}}, \bibinfo {author} {\bibfnamefont {R.}~\bibnamefont {Kullock}},
  \bibinfo {author} {\bibfnamefont {J.}~\bibnamefont {Prangsma}}, \bibinfo
  {author} {\bibfnamefont {M.}~\bibnamefont {Emmerling}}, \bibinfo {author}
  {\bibfnamefont {M.}~\bibnamefont {Kamp}}, \ and\ \bibinfo {author}
  {\bibfnamefont {B.}~\bibnamefont {Hecht}},\ }\href {\doibase
  10.1038/nphoton.2015.141} {\bibfield  {journal} {\bibinfo  {journal} {Nature
  Photon}\ }\textbf {\bibinfo {volume} {9}},\ \bibinfo {pages} {582} (\bibinfo
  {year} {2015})}\BibitemShut {NoStop}%
\bibitem [{\citenamefont {Johnson}\ and\ \citenamefont
  {Christy}(1972)}]{johnson_optical_1972}%
  \BibitemOpen
  \bibfield  {author} {\bibinfo {author} {\bibfnamefont {P.~B.}\ \bibnamefont
  {Johnson}}\ and\ \bibinfo {author} {\bibfnamefont {R.~W.}\ \bibnamefont
  {Christy}},\ }\href {\doibase 10.1103/PhysRevB.6.4370} {\bibfield  {journal}
  {\bibinfo  {journal} {Phys. Rev. B}\ }\textbf {\bibinfo {volume} {6}},\
  \bibinfo {pages} {4370} (\bibinfo {year} {1972})}\BibitemShut {NoStop}%
\end{thebibliography}%

\onecolumngrid
\clearpage
\begin{center}
\textbf{\large Supplementary Information: Tuning the Optical Properties of a MoSe$_2$ Monolayer Using Nanoscale Plasmonic Antennas}
\end{center}

\setcounter{equation}{0}
\setcounter{figure}{0}
\setcounter{table}{0}
\renewcommand{\theequation}{S\arabic{equation}}
\renewcommand{\thefigure}{S\arabic{figure}}
\renewcommand{\bibnumfmt}[1]{[S#1]}
\renewcommand{\citenumfont}[1]{S#1}

\section{S1. Plasmonic Dipole Nanoantenna as the Nanophotonic Resonator}

\begin{figure}[!ht]
\includegraphics{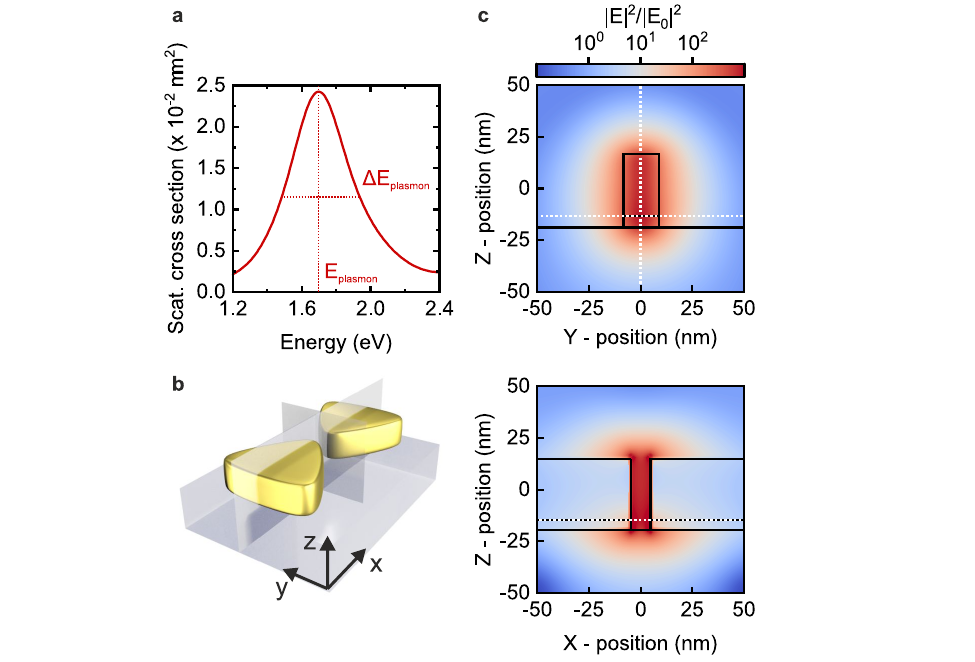}
\caption{\textbf{Quality factor $Q$ and mode volume $V_\mathrm{m}$ of the bowtie dipole plasmonic nanoantenna.} \textbf{(a)} Scattering cross section of the $d_{\mathrm{size}} = \SI{90}{nm}$ nanoantenna with feed gap size $d_{\mathrm{gap}} = \SI{10}{nm}$. The quality factor is evaluated as $Q = E_{\mathrm{plasmon}}/\Delta E_{\mathrm{plasmon}}$ and has typical values of $Q = 4-5$. \textbf{(b)} Schematic representation of a dipole nanoantenna with highlighted XZ- and YZ-planes that cross in the middle of the feed gap. \textbf{(c)} Spatial distribution of light intensity enhancement within the feed gap in YZ-plane (top) and XZ-plane (bottom). Substrate-air boundary and nanoantenna tip are indicated with black lines. Maximum enhancement occurs at the height of the Ti-Au interface close to the substrate, as indicated with white dotted lines.}
\label{fig:S1}
\end{figure}

The quality factor of a resonator can be determined via $Q = E_{\mathrm{plasmon}}/\Delta E_{\mathrm{plasmon}}$, where $E_{\mathrm{plasmon}}$ is the energy of the plasmonic resonance and  $\Delta E_{\mathrm{plasmon}}$ is its full width at half maximum (FWHM). This can be calculated by performing finite-difference time-domain (FDTD) simulations for the scattering cross section (Figure \ref{fig:S1}a). $Q$ is a measure for how well the resonator preserves energy over time. Nanoresonators used in this work typically have $Q=4-5$. The mode volume $V_{\mathrm{m}}$ can be calculated via: $V_{\mathrm{m}}=\frac{\int \epsilon E^2 \mathrm{d}V}{\max(\epsilon E^2)}$, where $\epsilon$ is the dielectric permittivity at the specific point, and $E$ is the electric field. The integral is performed over the feed gap volume. To reduce the computation time, it is also possible to estimate the mode volume by multiplying the lengths of the three Cartesian directions (white dotted lines in Figure \ref{fig:S1}c), where the field is higher than maximal intensity value divided by $e$. The estimated mode volume is $V_{\mathrm{m}}\sim \SI{2000}{\nano\meter^3}$.

\section{S2. Nominally Fixed Fabrication Parameters}

The plasmonic properties of the nanoantenna depend on the overall shape, size and feed gap size of the nanoantenna, the thickness of the gold, and the tip radius. Moreover, the resonance frequency is sensitive to the materials used for the nanoantenna and substrate materials, the presence of an adhesion layer, and the overall dielectric environment. 

\begin{figure}[!ht]
\includegraphics[width=1\textwidth]{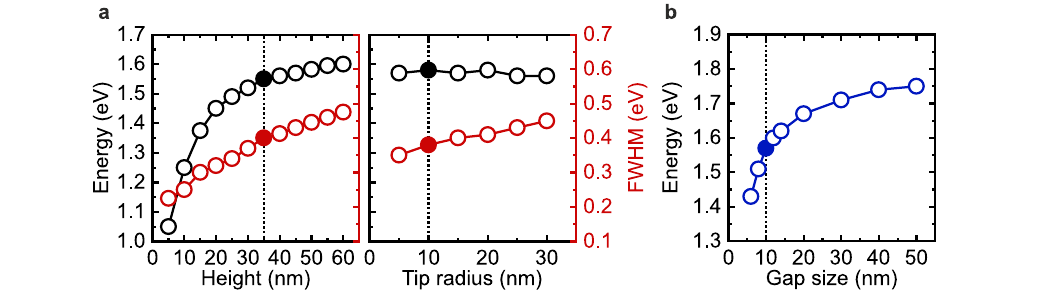}
\caption{\textbf{Tailoring hybrid TMD-gold nanostructures via FDTD calculations.} (\textbf{a}) Resonance energy (black) and FWHM (red) dependence on the nanoantenna height of the gold part (left) and tip radius (right). The optimal values are indicated with filled data points. \textbf{(b)} Resonance energy dependance on the feed gap size. The optimal value is denoted with the filled blue data point at $d_{\mathrm{gap}} = \SI{10}{\nano\meter}$.}
\label{fig:S2}
\end{figure}

By increasing the nanoantenna height (height of titanium is fixed at 5 nm, while the height of gold is changed), the plasmonic resonance energy and the linewidth increase as presented in Figure \ref{fig:S2}a on the left, obtained from FDTD simulations. The rate of change of the resonant energy slows down going from thinner to thicker gold nanoantennas, resulting in a plateau after 35 nm. As we increase the height, plasmons that live at different interfaces get decoupled, leading to higher energies. On the other hand, the FWHM is linearly increasing, since the damping is proportional to the volume of the nanoantenna. As a compromise, we choose the point where the rate of change of plasmonic resonances is rather small with the smallest possible FWHM, denoted with filled colored points.

The tip radius is dictated by fabrication resolution. Better resolution ensures a slightly higher quality of nanoantennas which is determined by the FWHM since the resonance energy does not change significantly, as shown in Figure \ref{fig:S2}a on the right. Here, we mark the points at 10 nm, since that is the estimated value for the tip radius limited by fabrication resolution.

In Figure \ref{fig:S2}b, we show the plasmonic resonance energy dependence on feed gap sizes $d_{\mathrm{gap}}$, from where we can demonstrate that the interaction between two nanoantenna arms has a $\sim d_{\mathrm{gap}}^{-3}$ dependence in accordance with the theory. The same results are obtained if we consider two nanoantenna arms as two electric dipoles interacting with each other, thus introducing the energy shift upon the change of distance. To obtain superior plasmonic cavities and reach large values of the $Q/V_{\mathrm{m}}$ ratio, we aimed to make small feed gaps between the nanoantenna arms, which would result in small mode volumes. We fixed the feed gap to be 10 nm, which is due to the estimated fabrication limit.

\section{S3. Reliable Design Using FDTD Simulations}

\begin{figure}[!t]
\includegraphics[width=1\textwidth]{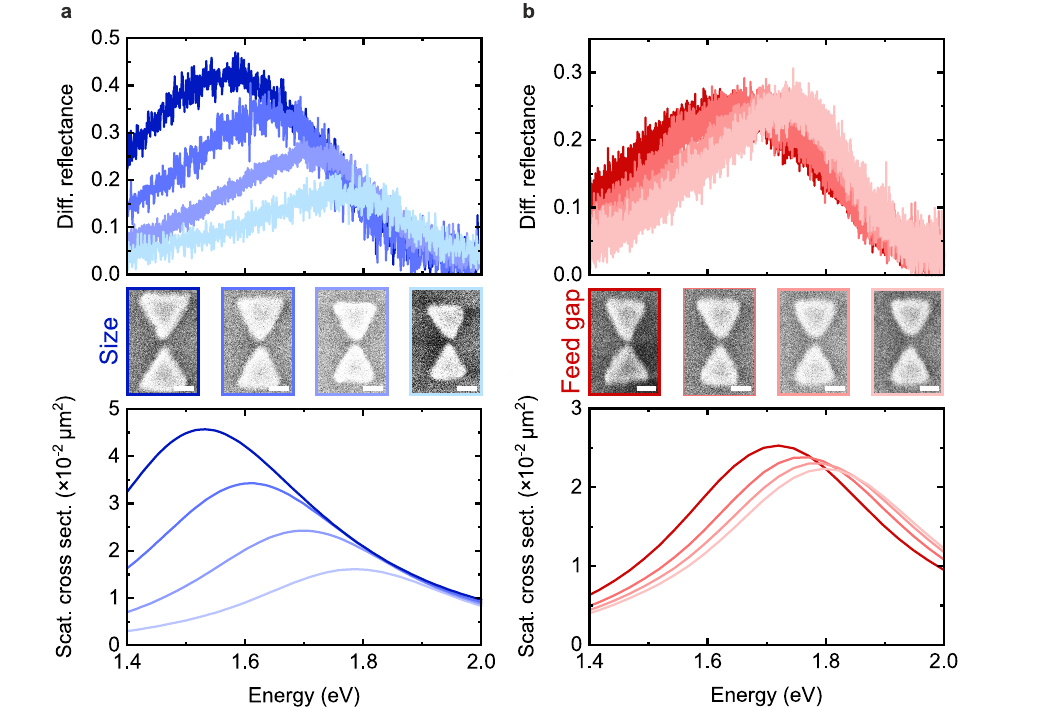}
\caption{\textbf{Differential reflectance dependence on nanoantenna arm size and feed gap size including the corresponding scattering cross sections obtained from theoretical predictions.} \textbf{(a)} Differential refectance spectra of nanoantennas with different arm size $d_{\mathrm{size}}$  = 110 nm; 100 nm; 90n m; 80 nm (upper plot) (SEM image insets highlighted with corresponding colors) and the corresponding theoretically predicted scattering cross sections (lower plot). \textbf{(b)} Differential reflectance spectra of nanoantennas of different feed gap sizes  $d_{\mathrm{gap}}$ = 8 nm; 10 nm; 12 nm; 14 nm (upper plot) (SEM image insets highlighted with corresponding colors) and the corresponding theoretically predicted scattering cross sections (lower plot). All scale bars are 50 nm.}
\label{fig:S3}
\end{figure}

The upper plot in Figure \ref{fig:S3}a shows a typical series of differential reflectance spectra recorded for nanoantennas with different arm sizes ranging from $d_{\mathrm{size}}$ = 110 nm (dark blue) down to $d_{\mathrm{size}}$ = 80 nm (light blue) in steps of 10 nm, as shown in the corresponding scanning electron microscope (SEM) image insets, while the feed gap size is fixed at 10 nm. As the size of an nanoantenna decreases, a shift of the plasmonic resonance to higher energies is observed. This can be intuitively understood by realizing that upon decreasing the size, the distance between surface charges at the opposite interface of the metallic nanoparticle decreases, which leads to a higher restoring force, hence a higher resonant energy. In addition, it can be seen that the linewidth decreases with decreasing size. When lowering the material volume, radiative damping is decreased, and the lifetime of the LSPP is increased. In addition, we observe that the amplitude contrast decreases with decreasing sizes since light scatters less for smaller nanoantennas. This prevented us from characterizing nanoantennas smaller than 60 nm due to the unfavorable signal-to-noise ratio. The lower plot shows the corresponding simulated scattering cross sections of plasmonic nanoantennas with respect to energy. Comparing the two plots, we observe excellent quantitative agreement between measurements and theoretical predictions with an offset of less than 0.1 eV. This is a good indication that we can rely on our theoretical treatment for the purpose of choosing the geometrical parameters for nanoantenna tailoring with the reproducible and controllable fabrication method. 

Another parameter which can be varied is the size of the nanoantenna feed gap. In Figure \ref{fig:S3}b, differential reflectance spectra of nanoantennas of different feed gap sizes are provided (upper plot), alongside with the corresponding theoretical prediction (lower plot). The feed gap size varies from  $d_{\mathrm{gap}}$ = 8 nm (dark red) to  $d_{\mathrm{gap}}$ = 14 nm (light red) in steps of 2 nm, as shown in the corresponding SEM image insets, while the arm size is fixed to 90 nm. A plasmonic resonance shift of 0.1 eV is observed when going from  $d_{\mathrm{gap}}$ = 8 nm to  $d_{\mathrm{gap}}$ = 14 nm. This trend can be explained by considering two nanoantenna arms as two electric dipoles. As the distance between the two is increased, the coupling weakens, thus the resonant energy is increased. As in the case of size variation, the theoretical predictions match the experiment quantitatively with an offset of 0.1 eV to higher energies.

\section{S4. Fabrication of Plasmonic Dipole Nanoantennas}

\begin{figure}[!t]
\includegraphics[width=1\textwidth]{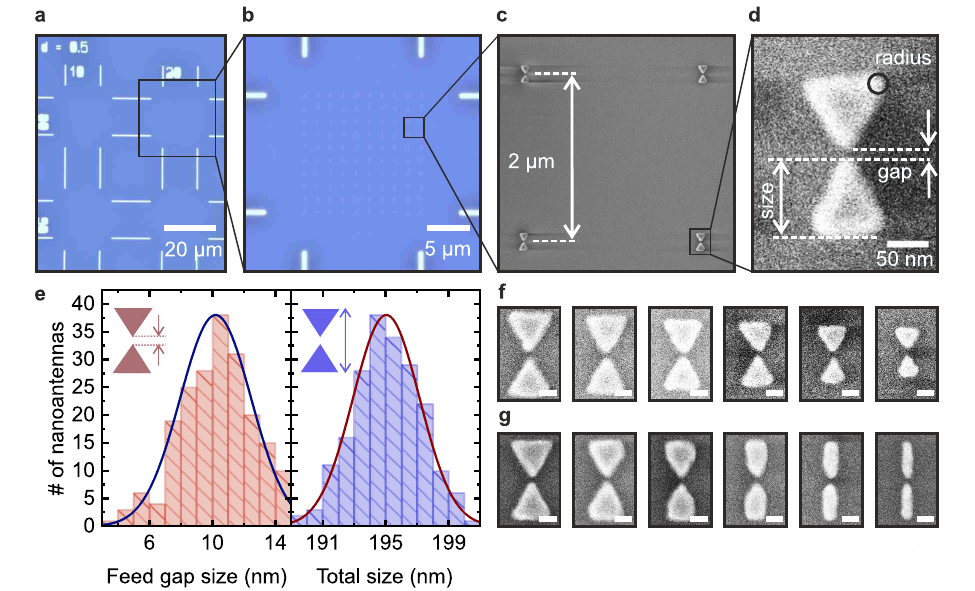}
\caption{\textbf{Typical plasmonic bowtie dipole nanoantenna sample.} \textbf{(a)} Microscope image showing the sample layout. Dose and size are varied over different fields.
\textbf{(b)} Microscope image of the central region of a writing field. \textbf{(c)} SEM image of neighboring nanoantennas spaced by \SI{2}{\micro\meter}. \textbf{(d)} SEM image of a single nanoantenna, where size, feed gap size, and radius of curvature are indicated. \textbf{(e)} Distribution of feed gap sizes $d_{\mathrm{gap}}=\SI{10\pm2}{\nano\meter}$ (blue) and total sizes $2d_{\mathrm{size}} + d_{\mathrm{gap}}=\SI{195\pm2}{\nano\meter}$ (red) for 200 nanoantennas. \textbf{(f)} SEM images of plasmonic nanoantennas of nominal sizes ranging from 110 nm down to 60 nm in intervals of 10 nm. \textbf{(g)} SEM images of truncated nanoantennas ranging from triangular shape to rod-like shape with a width of 20 nm. All scale bars in \textbf{(f)} and \textbf{(g)} are 50 nm.}
\label{fig:S4}
\end{figure}

A microscope image of a typical sample layout is presented in Figure \ref{fig:S4}a. We varied two parameters, electron-beam dose and nanoantenna size, while the nominal values for feed gap size (10 nm), radius of curvature (10 nm) and gold height (35 nm) remained fixed. To achieve the high precision needed for small feed gap sizes, we were using high doses ranging from $\SI{400}{\micro\coulomb/\centi\meter^2}$ to $\SI{800}{\micro\coulomb/\centi\meter^2}$. Figure \ref{fig:S4}b shows the central region of a writing field, which contains 100 nanoantennas for the purpose of better statistics. Nanontennas were positioned \SI{2}{\micro\meter} from each other to avoid coupling of neighbouring nanoantennas, as shown by the SEM image in Figure \ref{fig:S4}c. This enabled us to address a single bow-tie nanoantenna in our optical experiments. In  \ref{fig:S4}d, a typical bow-tie nanoantenna is shown with 80 nm size, 10 nm feed gap size and 10 nm radius of curvature, imaged with SEM. The two histograms in Figure \ref{fig:S4}e and Figure \ref{fig:S4}f, showing the feed gap and nanoantenna sizes of 200 nanoantennas, give us an insight into the reproducibility and correctness with respect to the nominal sizes of nanoantennas. We were able to reproducibly obtain nanoantennas with feed gaps of $d_{\mathrm{gap}}=\SI{10\pm2}{\nano\meter}$, for nominal feed gap sizes of 10 nm. We were able to reach sub-10nm feed gaps, achieving state-of-the-art fabrication resolution. Measuring the distance from one end of nanoantenna to the other (therefore, counting the total length of 2 nanoantenna arms and a feed gap) we obtained values of  $2d_{\mathrm{size}} + d_{\mathrm{gap}} = \SI{195\pm2}{\nano\meter}$ for nominal values of \SI{190}{\nano\meter}.

Tailoring the size of nanoantennas enables plasmon resonance tuning to desired wavelengths. We were able to produce nanoantennas as small as  $d_{\mathrm{size}}=\SI{60}{\nano\meter}$ (shown in Figure \ref{fig:S4}g), which corresponds to plasmonic resonances centered at \SI{1.95}{\electronvolt}. As the size increases, the resonance shifts to the red, meaning that we covered a wide spectral range from the near-infrared regime to \SI{1.95}{\electronvolt}. To increase the $Q$ factor of the nanoantennas, we designed the new nanoantennas by gradually truncating bowtie nanoantennas to obtain rod-like nanoantennas with sharp tips and with widths as small as $w=\SI{20}{\nano\meter}$, as in Figure \ref{fig:S4}f. However, these nanoantennas were not used in our experiments, due to their decreased reflectance compared to bow-tie nanoantennas, owing to their smaller gold volume and causing difficulties in the form of low signal in the differential reflectance measurements. Therefore, a bowtie geometry was used for producing plasmonic nanoantennas, which we later coupled to a TMD monolayer.

\section{S5. Fitting Parameters}

\begin{figure}[!t]
\includegraphics[width=1\textwidth]{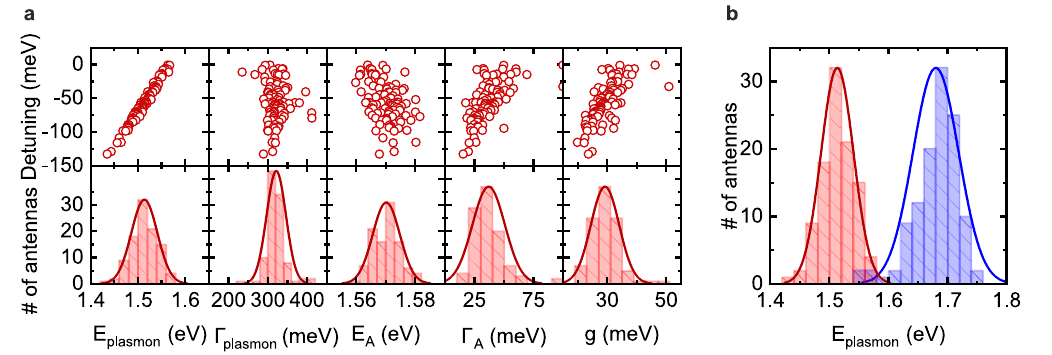}
\caption{\textbf{Fitting to the COM.} \textbf{(a)} Distribution and detuning dependence of resonance plasmon energy $E_{\mathrm{plasmon}}$, plasmon linewidth $\Gamma_{\mathrm{plasmon}}$, exciton energy $E_\mathrm{A}$, exciton linewidth $\Gamma_\mathrm{A}$, and coupling constant $g$. \textbf{(b)} Plasmonic resonance energies of $\sim$120 bare nanoantennas (blue) and $\sim$100 TMD-gold hybrid nanostructures (red).}
\label{fig:S5}
\end{figure}

In Figure \ref{fig:S5}a we show the correlation with the detuning (top) and the distribution (bottom) for each of the parameters of the fit. The plasmon energy takes values $E_{\mathrm{plasmon}} = \SI{1.51\pm0.03}{\electronvolt}$. The correlation of detuning and energy simply originates from the definition of detuning. The plasmon linewidths take values of  $\Gamma_{\mathrm{plasmon}} = \SI{320\pm25}{\milli\electronvolt}$ and do not show prominent dependence on detuning. The same is the case for exciton energies, which are narrowly distributed around  $E_{\mathrm{A}} = \SI{1.570\pm0.005}{\electronvolt}$. The exciton linewidths exhibit broadening as zero detuning is approached and have values of  $\Gamma_{\mathrm{A}} = \SI{37.0\pm1.3}{\milli\electronvolt}$. Finally, the coupling constant takes values $g = \SI{29\pm5}{\milli\electronvolt}$. 

In Figure \ref{fig:S5}b we plot the distribution of the plasmonic resonances for 115 bare nanoantennas (blue data) and 102 TMD-gold hybrid nanostructures (red data), extracted from the fits. The plasmonic resonance shifts by approximately \SI{170}{\milli\electronvolt} to lower energies. To deterministically engineer nanoantennas for coupling to A excitons, this shift has to be taken into account. Although theoretical calculations predicted the shift, it was underestimated in practical realization. It is shown in Figure \ref{fig:S5}b that the new resonances are distributed away from the $A$ exciton energy, with all of the nanoantennas negatively detuned from \SI{1.57}{\electronvolt}. To perfectly position the plasmon resonances in the red histogram (i.e. to perfectly tune TMD-gold hybrid nanostructures) to the position of the $A$ exciton, smaller nanoantennas are needed. This is a very challenging task on both the fabrication and the optical characterization front. Due to the low volume of the structure, the signal-to-noise ratio decreases in differential reflectance measurements. This analysis and a need to reach perfect plasmon-exciton matching, motivated us to produce a new sample, which will be presented in the next section.

\section{S6. Sample with Positive Detuning}

\begin{figure}[!ht]
\includegraphics[width=0.7\textwidth]{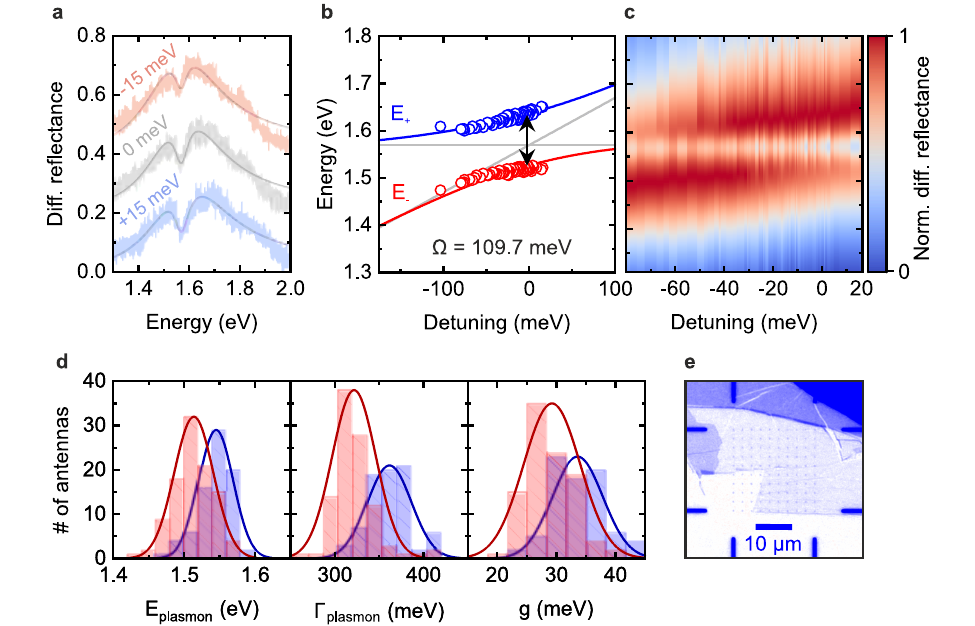}
\caption{\textbf{Reaching positive detuning with $\mathrm{MoSe_2}$-coupled dipole nanoantennas containing nanoantennas of $d_{\mathrm{size}} = \SI{80}{\nano\meter}$ and $d_{\mathrm{gap}} = \SI{10}{\nano\meter}$.} \textbf{(a)} Differential reflectance spectra and their fits for detunings around zero. \textbf{(b)} High- and low- energy branches dependence on detuning showing anti-crossing behavior. \textbf{(c)} Normalized fits to differential reflectance spectra of 78 TMD-gold hybrid nanostructures. \textbf{(d)} Distribution of plasmon energy, plasmon linewidth and coupling constant for samples with \SI{90}{nm} sized nanoantennas (red) and \SI{80}{nm} sized nanoantennas (blue). \textbf{(e)} Red channel contrast image of the nanoantenna field after the monolayer transfer.}
\label{fig:S6}
\end{figure}

After characterization of 78 nanoantennas, we treated the data in a similar fashion as for the previous sample. In Figure \ref{fig:S6}a, we show raw spectra and corresponding COM fits in the vicinity of zero detuning. Unlike the previous time we fixed the exciton energy to $E_\mathrm{A} = \SI{1.57}{\electronvolt}$ and exciton linewidth to $\Gamma_\mathrm{A} = \SI{0.04}{\electronvolt}$, to obtain better fits. These parameters are extracted from differential reflectance measurements on the bare flake. Analysis of our data shows that the new sample exhibits a higher splitting at zero detuning and has a value of $\Omega = 2g|_{\delta = 0} = \SI{110}{\milli\electronvolt}$. Comparing $g|_{\delta = 0} = \SI{55}{\milli\electronvolt}$ and the critical coupling constant $g_c = \SI{100}{\milli\electronvolt}$, we again conclude that our system is operating in the Fano regime. Figure \ref{fig:S6}b shows high and low energy branches which demonstrate clear anti-crossing behavior in contrast to the uncoupled plasmon and exciton lines denoted with gray. Normalized fits to the experimental data obtained from 78 different TMD-gold nanostructures are shown in Figure \ref{fig:S6}c, where the positively detuned spectra, which were not present in the previous sample, are now present in the detuning range from 0 meV to 20 meV. Finally, we compare different parameters and their behavior for the two samples in Figure \ref{fig:S6}d. The plasmon energies are shifting to higher energies since the new sample consists of smaller nanoantennas than the previous one and have values $E_{\mathrm{plasmon}} = 1.545\pm\SI{0.025}{\electronvolt}$. The same trend is assigned to the plasmon linewidth with a value of $\Gamma_{\mathrm{plasmon}} = 360\pm\SI{25}{\milli\electronvolt}$. This goes against our intuition since we expect narrower linewidths for smaller nanoantennas. The reason is lower fabrication quality, resulting in nanoantennas with lower $Q$ factors. The coupling constant $g$ in the new sample is on average increased for TMD-gold nanostructures, and has a value of $g = 34\pm\SI{4}{\milli\electronvolt}$.
The contrast micrograph of the sample is represented in Figure \ref{fig:S6}e. 

\section{S7. AFM Topography - Appendix}

\begin{figure}[h]
\includegraphics[width=0.3\textwidth]{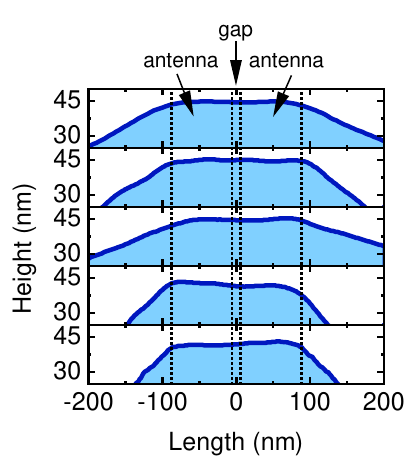}
\caption{\textbf{AFM profile along the long axis of five different hybrid TMD-antenna structures.} The monolayer is flat in the sensitive region at the gap and remains that way conforming to the nanoantenna surface. Outside of the nanoantenna region, the flake is suspended under different angles.}
\label{fig:S7}
\end{figure}

\section{S8. Large Redshift upon Monolayer Transfer onto Nanoantennas}

\begin{figure}[h]
\includegraphics{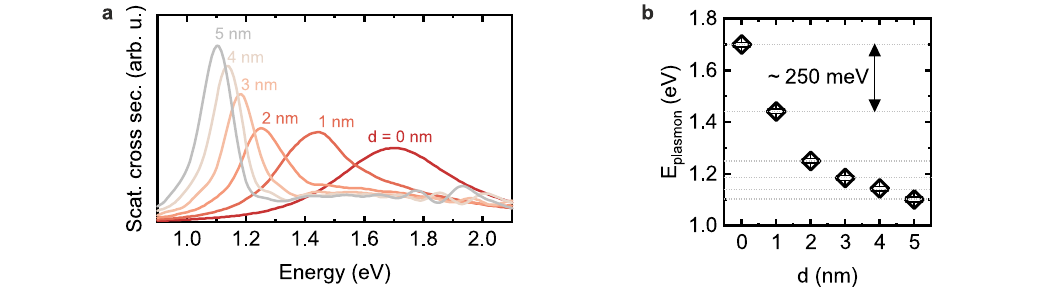}
\caption{\textbf{Redshift of the plasmonic resonance upon introduction of the high-refractive-index material in proximity of plasmonic hot-spot.} \textbf{(a)} Simulated scattering cross section spectra for different thicknesses $d$ of the high-refractive-index material mimicking TMD monolayer (excluding excitonic and non-local screening effects). \textbf{(b)} Plasmonic resonance energy as a function of the TMD-like slab of thickness $d$.}
\label{fig:S8}
\end{figure}

In our samples, a very small spatial change in the dielectric environment of the plasmonic resonator results in a big change of the plasmon energy. This can be explained by the following arguments. TMDs are materials with a relatively high refractive index $n$. In addition, it also matters where the material is positioned with respect to the hot-spot. Here, we have a high refractive index material that is in close proximity to the complete top of the structure and is highly influencing the mode shape that is accommodated in the feed gap. We have performed FDTD calculations where we positioned a thin slab of the refractive index $n=4.5$ (that mimics a TMD monolayer excluding the excitonic and non-local screening effects) on top of a golden nanoantenna. Then, we vary the slab’s thickness from 0 nm to 5 nm in increments of 1 nm and record the scattering cross section, depicted in Figure \ref{fig:S8}a. Upon the introduction of a 1 nm thin high refractive index material, the plasmonic resonance exhibits a large redshift of 250 meV, which can be clearly seen in Figure \ref{fig:S8}b. This aligns well with the shift we observe (200 meV for 0.65 nm). The resonance shifts even further by increasing thickness, but with a slower rate as the thickness goes up.

\end{document}